\documentclass[reprint,aps,pre]{revtex4-1}

\usepackage{graphicx}
\usepackage{bm}
\usepackage{epstopdf}
\usepackage{amsmath,amsfonts,paralist}
\usepackage{ amssymb }
\usepackage{appendix}
\usepackage{mathbbol}
\let\a=\alpha     \let\d=\delta \let\e=\varepsilon
     \let\l=\lambda
                
\let\s=\sigma     
   
\let\G=\Gamma

\let\la=\langle
\let\ra=\rangle

\def\to{\rightarrow}
\def\la{\left\langle}
\def\ra{\right\rangle}

\newcommand{\wh}{\widehat}

\newcommand{\Tr}{\text{Tr}}
\newcommand\norm[1]{\left\lVert#1\right\rVert}

\begin{document}

\title[Learning Maximum Entropy Models from finite size datasets]{Learning Maximum Entropy Models from finite size datasets: a fast Data-Driven algorithm allows sampling from the posterior distribution}

\author{Ulisse Ferrari}

\affiliation{Sorbonne Universit\'es, UPMC Univ Paris 06, INSERM, CNRS, Institut de la Vision, 17 rue Moreau, 75012 Paris, France.}

\begin{abstract}
Maximum entropy models provide the least constrained probability distributions that reproduce statistical properties of experimental datasets.
In this work we characterize the learning dynamics that maximizes the log-likelihood in the case of large but finite datasets.
We first show how the steepest descent dynamics is not optimal as it is slowed down by the inhomogeneous curvature of the model parameters space.  
We then provide a way for rectifying this space which relies only on dataset properties and does not require large computational efforts.
We conclude  by solving the long-time limit of the parameters dynamics including the randomness generated by the systematic use of Gibbs sampling.
In this stochastic framework, rather than converging to a fixed point, the dynamics reaches a stationary distribution, which for the rectified dynamics reproduces the posterior distribution of the parameters.

We sum up all these insights in a ``rectified'' \textit{Data-Driven} algorithm that is fast and by sampling from the parameters posterior avoids both under- and over-fitting along all the directions of the parameters space.
Through the learning of pairwise Ising models from the recording of a large population of retina neurons, we show how our algorithm outperforms the steepest descent method.
\end{abstract}


\maketitle

Nowadays scientists from many different disciplines face the problem of understanding and characterizing the behavior of large  multi-units complex systems with strong correlations \cite{Buzaki04,Phillips04,Schneidman06,Peyrache09,Weigt09,Cocco13,Bialek14,Santolini14}.
Statistical Inference tackles these problems by inferring parameters of a chosen, context inspired,  probability distributions to obtain models reproducing the system behavior.
The basic strategy consists in choosing a model family  described by a set of parameters and tune them to reproduce the dataset properties.

However, in many cases, due to a substantial unawareness of the system properties or to avoid biasing the results,  hypotheses on the distribution functional form cannot be suggested nor trusted.
To overcome this issue, the Maximum Entropy (MaxEnt) Principle\cite{Jaynes82} suggests to search for the probability distribution with the largest entropy between those satisfying a set of constraints, which force the model distribution to reproduce the experimental averages of a list of observables.
For systems of binary variables a common and fruitful choice is to constraint the model distribution to reproduce the experimental single and pairwise correlations.
This choice has been successfully  applied in system neuroscience \cite{Schneidman06,Cocco09,Hamilton13,Mora15,Tavoni14},
gene regulation \cite{Lezon06}, fitness estimation \cite{Ferguson13,Mann14} and many others fields of science.
Moreover other possibilities with different observable lists have been also investigated \cite{Ganmor11,Tkacik14}, suggesting that depending on the context a careful choice of the observables can improve the inference accuracy and predicting power.

However, once the MaxEnt problem is posed as an inference task, finding its solution can be very hard. 
For large system size, in fact, the inference problem cannot be solved analytically and specifically devoted algorithms are required.
The most known and widely used algorithm \cite{Ackley85} was introduced in the eighties and later developed \cite{Hinton02,Broderik07}.
Other approaches include Selective Cluster Expansions \cite{Cocco11,Barton13}, Minimum Probability Flow \cite{Sohl-Dickstein11} and several approximation schemes \cite{Kappen97,Tanaka98,Aurell12,Ricci12,Jacquin15}.

In this paper we develop further the approach of \cite{Ackley85}.
Our results are based on an analysis of the geometrical structure of the model parameters space.
For the MaxEnt models inference this space is shown to benefit of peculiar properties, which allow us to introduce a novel \textit{quasi-Newton} method, the \textit{Data-Driven} algorithm, and to completely characterize its long-time learning dynamics.
As it is affected by the randomness of Monta Carlo estimates, the parameters dynamics is stochastic and eventually converges to a stationary distribution around the log-likelihood maximum.
The presented method takes advantage of this randomness and shapes the stationary distribution to reproduce the Bayesian posterior distribution and thus to sample from it.
This last feature allows to avoid overfitting and endorses our algorithm to be well suited for dataset with highly in-homogeneous noise.
We conclude with a test on biological data.

\section{Maximum Entropy Models} 
\label{sect::MaxEnt}

Systems of interest for MaxEnt approach are composed by $N$ units that show a stochastic and coupled behavior.  
In order to be concrete, in this work we focus on binary, $0/1$, units, but most of the results can be directly generalized to multiple-state variables, as Potts or Poisson models, or even continuous ones. 
In this framework, datasets are composed by $B$ independent measurements of the synchronous state of the $N$ system units: $\{ \{  \s_i(b) \}_{i=1}^N\}_{b=1}^B$.
As an example, for a binned spike trains recording of several neurons, $\s_i(b)=1$ may represent the activity (spike) of the i-th neuron in the b-th time-bin and $\s_i(b)=0$ its silence. 

The first, crucial, step in the MaxEnt analysis is the choice of a set of observables.
Observables are generic functions of the system units that should be chosen in order to catch  the way system components interact.
Strictly speaking, if some statistical feature is relevant for the system behavior, the corresponding observable should be included in order to force the model to reproduce that feature.
On the other way round, if some feature is not essential, by excluding the corresponding observable, the model will adapt its behavior consistently with the other imposed features.
Technically speaking the observables will be the sufficient statistics of the model probability distribution.
As an example, the most common choice in the literature considers all the single variable terms ($\s_i$) and pairwise products ($\s_i\s_j$) and it leads to the construction of the well know Disorder Pairwise Ising model \cite{Schneidman06}.
This particular choice allows to take into account the pairwise correlations between the units and allows the model to adapt higher order statistics consistently.


Moreover, a carefully choice of the observable should take into account the quality of the dataset.
Noisy observables  for which the average is not significantly estimated will induce data overfitting with the risk of strongly reducing the model prediction power.
From this point of view and overloading the jargon of Bayesian Inference, the observables choice represents a sort of prior term (see sect. \ref{sect::Bayes} for details).

For the sake of generality, we like to present the results for an arbitrary observables choice and with this aim in mind we introduce a generic vector of observables ${\bf \Sigma}({\bf \s }) \equiv \{ \Sigma_a ( {\bf \s }) \}_{a=1}^D$, functions of the system units.
The choice of the observables vector completely determines the functional form of the MaxEnt model family \cite{Jaynes82,Cocco11}.
In fact, by searching for the probability distribution which has the largest entropy among those that reproduce the observables averages, we obtain (see, for example \cite{Barton13}, for the whole functional calculation): 
\begin{equation}
P_{\bf X}( {\bf \s}) = \exp \{ {\bf \Sigma(\sigma) \cdot  X }\}  /  Z[ {\bf X} ]
\label{boltzDistr}
\end{equation}
where  $Z[{\bf X}]$ is a normalization constant, ${\bf X}$ is a $D$ dimensional fields vector, namely the model parameters, conjugated to the observable vector ${\bf \Sigma}$ and ${\bf \Sigma(\sigma) \cdot  X } = \sum_a  \Sigma_a(\sigma)  X_a$ is the scalar product in Euclidean space\footnote{In order to lighten the notation we do not distinguish between column or row vector and we avoid any transpose symbol.}. 

As the distribution (\ref{boltzDistr}) measures the probability of each possible system configurations, we can compute its dataset (log-)likelihood:
\begin{equation}
 l\big(~\text{data}~ \big| ~{\bf X} ~\big) \equiv \sum_{b=1}^B \ln P_{\bf X}\big( {\bf \s}(b)\big).
\end{equation}
Even if later we will consider posterior sampling, see sect. \ref{sect::Bayes}, for the moment we restrict the inference task to the search for the set of fields ${\bf X^*}$ that maximizes the log-likelihood:
\begin{equation}
{\bf X^*} \equiv \arg \max_{\bf X} \Big[ ~l[{\bf X}] ~\Big] \label{maximization}
\end{equation}
where
\begin{equation}
l[{\bf X}] = B \Big(~ {\bf X \cdot \overline{P} }  -  \ln Z[{\bf X}] ~\Big),
\label{logL}
\end{equation}
and
\begin{equation}
\overline{\bf P}  \equiv  \overline{\bf \Sigma } \equiv \frac{1}{B} \sum_{b=1}^B {\bf \Sigma}(\sigma(b)~) 
\end{equation}
are the experimental averages of the observables.
In fact, finding the fields values ${\bf X^*}$ solving
\begin{equation}
\nabla_a  l[{\bf X^*}]  = B \Big(\overline{P_a} - Q_a[{\bf X^*}]\Big) = 0~,
\label{gradient}
\end{equation}
where $\nabla_a \equiv d/ dX_a$, is equivalent to enforce the constraints:
\begin{equation}
{\bf Q[X^*]} =  \overline{\bf P} 
\label{inferenceEquation}
\end{equation}
where
\begin{equation}
{\bf Q[X]} \equiv \langle {\bf \Sigma } \rangle_{\bf X} \equiv \Tr_\sigma \big[ {\bf \Sigma }(\sigma) P_{\bf X}( {\bf \s})\big]~. \label{modelAverage} 
\end{equation}
are the model averages of the observables.
Here and in the following $\langle \dots \rangle_{\bf X}$ means average over the model distribution (\ref{boltzDistr}) with fields ${\bf X}$ and $\overline{(\dots)}$ always refers to dataset averages.  $\Tr_\sigma$ means summation over all possible system configurations. 

For most of the reasonable observables choice and in case of good data quality, the solution of the maximization problem (\ref{maximization}) exists and is unique.
However for sake of completeness, in app. \ref{app::Existence} we discuss  when and how multiple solutions could arise. 

\subsection{The geometry of the ${\bf X}$- space}
\label{geometry}

In the following sections we will deal with the non trivial geometry embedding the fields space, the ${\bf X}$-space.
This geometry is described by three matrices: the (negative) log-likelihood Hessian $\mathcal{H}[{\bf X}]$, the Fisher matrix $I[{\bf X}]$ and the model susceptibility matrix $\chi[{\bf X}]$
\begin{eqnarray}
\mathcal{H}_{ab}[{\bf X}] &\equiv& -\nabla_a \nabla_b  l\big[{\bf X}\big] / B  \label{hessian}  ~,\\
I_{ab}[{\bf X}] &\equiv& \big\langle \nabla_a   \ln P_{\bf X}\big( {\bf \s}\big) ~  \nabla_b  \ln P_{\bf X}\big( {\bf \s}\big)   \big\rangle_{\bf X} \label{fisher} ~, \\
\chi_{ab}[{\bf X}] &\equiv&  \big\langle  \Sigma_a  \Sigma_b  \big\rangle_{\bf X} - \big\langle \Sigma_a \big\rangle_{\bf X} \big\langle \Sigma_b \big\rangle_{\bf X}   \label{chi} ~.
\end{eqnarray}
$\mathcal{H}[{\bf X}]$ describes the concavity of the log-likelihood function, $I[{\bf X}]$ describes the covariance of the log-likelihood gradient, whereas $\chi[{\bf X}]$ that of the observables.
If in a general optimization problem these three matrices differ, in the the MaxEnt model inference they coincide.
In fact:
\begin{eqnarray}
I_{ab}[{\bf X}] &\equiv&~ \big\langle \nabla_a   \ln P_{\bf X}\big( {\bf \s}\big) ~  \nabla_b  \ln P_{\bf X}\big( {\bf \s}\big)   \big\rangle_{\bf X}  \\
 &=& -  \big\langle \nabla_a  \nabla_b  \ln P_{\bf X}\big( {\bf \s}\big)   \big\rangle_{\bf X}  \\
&=& -\nabla_a \nabla_b  l\big[{\bf X}\big]/B~ \equiv~ \mathcal{H}_{ab}[{\bf X}]
\end{eqnarray}
 where in the first equality we use a well know properties of the Fisher matrix and in the second the fact that the log-likelihood is linear in $ \overline{\bf P} $, the only data-dependent quantities.
Moreover:
\begin{eqnarray}
I_{ab}[{\bf X}]  &=& -  \big\langle \nabla_a  \nabla_b  \ln P_{\bf X}\big( {\bf \s}\big)   \big\rangle_{\bf X}  \\
&=& ~  \nabla_a  \nabla_b  \ln Z[{\bf X}]  \\
&=& ~ \big\langle  \Sigma_a  \Sigma_b  \big\rangle_{\bf X} - \big\langle \Sigma_a \big\rangle_{\bf X} \big\langle \Sigma_b \big\rangle_{\bf X} ~ \equiv ~\chi_{ab}[{\bf X}] ~.
\end{eqnarray}
Indeed:
\begin{equation}
\chi_{ab}[{\bf X}] ~ = ~ I_{ab}[{\bf X}] ~=~ \mathcal{H}_{ab}[{\bf X}]~.
\label{equalities}
\end{equation}

The equations (\ref{equalities}) are the keystones of this study.
They affect the geometry of the ${\bf X}$-space and thus the inference in a peculiar way.
In particular, the first equality allows us to introduce an efficient inference method, whereas the second allows us to completely characterize its long-time dynamics.

\subsection{A Bayesian Framework for the Maximum Entropy Models}
\label{sect::Bayes}

Until now, we posed the MaxEnt inference as a log-likelihood maximization problem, without considering that the finite size of the dataset could affect the estimate of the observables mean.
The error in these estimates will inevitably result in some uncertainty on the fields inference that has to be taken into account.
In fact, even if a carefully choice of the ${\bf \Sigma}$ avoids to include noisy observables, the system heterogeneity will results in a different precision in the fields estimation.

A Bayesian framework including prior and posterior distributions will exactly account for this uncertainty. 
In fact, through Bayesian inversion, we can compute the posterior distribution of the fields, $P^\text{Post}\big(~{\bf X}\big| \text{data}~ \big)$, from the prior distribution on the fields and the likelihood function, $P\big( \text{data} \big| {\bf X} \big)= \prod_b  P_{\bf X}\big( {\bf \s}(b)\big) $ :
\begin{equation}
P^\text{Post}\big(~ {\bf X} ~\big| ~\text{data} ~ \big) = \frac{  P\big(~ \text{data}~ \big|~ {\bf X}~ \big) ~ P^\text{Prior}\big(~{\bf X}~\big) }{ \text{Norm} }.
\label{posterior}
\end{equation}
where Norm is a ${\bf X}$ independent normalization constant.
The width of the posterior distribution around its maximum quantifies the intrinsic uncertainty on the fields ${\bf X}$ and can be used to test the robustness of the inference.
Explicitly, any scientific results based on a particular outcome of the fields inference should remain valid for all the fields sets with large posterior probability.
Indeed the possibility to sample from the posterior is a powerful tool to test the robustness of the system analysis.

For the prior distribution we have two possible approaches: either include a flat distribution that does not depend on the fields value, either include a probability distribution that reflects some \textit{a priori} knowledge on them. 
The first approach lets the dataset account for the whole uncertainly, and in case of very good dataset should be preferred.
However, in applications dealing with strongly undersampled data and/or when some \textit{a priori} knowledge is available, the second possibility has shown to be powerful.

In this work, as an example to show how to include a prior term, we focus on the $L2$-regularization of the form:
\begin{equation}
P^\text{Prior}_{L2}\big(~{\bf X}~\big) \equiv  \exp\left\{ - \frac{B}{2} {\bf X} \cdot \eta \cdot {\bf X}  \right\} \Big/ \left( \frac{2 \pi}{|\eta| B} \right)^\frac{D}{2}, \label{L2prior}
\end{equation}
where $\eta$ is an arbitrary $D$-dimensional positive definite square matrix and $|\eta|$ is its determinant.

\section{Vanilla Gradient, Newton Method and Data-Driven Algorithms}

Many of the algorithms suited for solving the MaxEnt inference task performs a dynamics in fields space that, starting from an initial condition, flows toward the maximum of the log-likelihood.
In many applications, in fact, the log-likelihood gradient is fast to compute or estimate and can be used to drive the dynamics to the maximum.
In this section we first review two of these approaches and then we propose the \textit{Data-Driven} algorithm, the focus of this work.

\subsection{Review: Vanilla Gradient and Newton Method algorithms}

Before introducing the method proposed in this work, we like to review two well known inference method.

Ackley, Hinton and Sejnowski \cite{Ackley85} posed the inference problem as a dynamical process ascending the log-likelihood function along the gradient direction:
\begin{equation}
{\bf  X}_{t+1} - {\bf X}_t \equiv {\bf \d X}^\text{VG}_t  = \alpha {\bf (\overline{P} - Q[{\bf X}_t ])},
\label{VanillaGradient}
\end{equation}
where $\a$ is a learning rate.
The algorithm works iteratively: at each time-step the computation of the model averages ${\bf Q[{\bf X}_t ]}$ allows to perform the fields update and to proceed towards the convergence, which is guaranteed for sufficiently small $\alpha$.
We call this approach  \textit{Vanilla (Standard) Gradient (VG)} algorithm. 

Although it follows the gradient, VG  will not go trough the shortest path even for arbitrary small $\alpha$\cite{Amari98}.
The reason lies in the geometrical structure induced by the curvature of the log-likelihood function, namely its Hessian, see eq. (\ref{hessian}).
To take into account this geometrical effect we can multiply the log-likelihood gradient by the inverse of the Hessian \cite{Press07} obtaining the well known Newton-Raphson method.
However as suggested by Amari\cite{Amari98b,Amari07}, in the curved manifold of the log-likelihood the natural local metric is the model Fisher matrix $I[{\bf X} ]$, see eq. (\ref{fisher}).
Indeed, in the geometry induced by $I[{\bf X} ]$, the steepest ascendant direction is  $ I^{-1} \nabla  l$, the contravariant form of the gradient (\ref{gradient}) \cite{Amari98}.
However for the MaxEnt models inference, Hessian and Fisher matrix coincides and so do Newton-Raphson and Natural gradient methods.
Consequently, we do not need to distinguish and  simply replace the VG update (\ref{VanillaGradient}) with 
\begin{equation}
{\bf \d X}^\text{NM}_t  = \alpha {\bf \chi}^{-1}[{\bf X}_t] \cdot (\overline{P} - Q[{\bf X}_t ])
\end{equation}
to obtain both the \textit{Newton Method (NM)} and the Natural gradient.
The positiveness of ${\bf \chi}$ ensures the convergence of this method at least for infinitesimally small $\a$  (see later).
Despite it is optimal, the Newton Method is slowed down by the time-consuming estimation and inversion of ${\bf \chi}[{\bf X}]$ at each update step. 
In the following we suggest a way to \textit{bypass} this problem.

\subsection{The Data-Driven Algorithm.}

As ${\bf \chi}[{\bf X}]$ depends only on model averages of observables products, we can approximate its value at the solution  ${\bf X = X^*}$ with the dataset configurations list: 
\begin{equation}
\chi_{ab}[{ \bf X^*}] \approx \overline{~ \chi_{ab}~} \equiv  \overline{ ~\Sigma_a ~  \Sigma_b~}- \overline{ ~\Sigma_a} ~\overline{ ~\Sigma_b}~.
\label{chiApprox}
\end{equation}
This matrix can be computed before running the inference dynamics and then used to evaluate the fields update.
The resulting \textit{Data-Driven} (DD) quasi-Newton  Method update rule reads:
\begin{equation}
 {\bf \d X }^\text{DD}_t = \alpha~ {\overline{{~\bf \chi}~}  }^{-1} \cdot  ( {\bf \overline{P} - Q} [{\bf X}_t ])~. \label{ApproximatedNewton}
\end{equation}

The quality of the approximation  ${\bf \chi}[{\bf X^* }] \approx \overline{~\bf\chi ~}$ depends on two main hypotheses: 
i) the ability of the MaxEnt model to reproduce dataset statistical properties beyond the mean of ${\bf \Sigma}$ and in particular the  experimental ${\bf \Sigma}$-covariance and ii) the good sampling quality of the experimental dataset.
The first hypothesis, however, was already partially assumed when it has been chosen the observables to reproduce and so the MaxEnt model to fit.
Otherwise stated, if the approximation is poor, it means that the chosen MaxEnt model was not a good choice.
The second hypothesis, instead, reflects the quality of the whole inference task.
In case of strong undersampling the inference problem, despite being mathematically well posed, is meaningless as the information encoded in the dataset does not support the fine tuning of the fields.
In sect. \ref{sect::Test} we will give a practical condition to test this second hypothesis.

The quality of the approximation  (\ref{chiApprox}) controls the speed up factors of the DD algorithm:
the better the approximation is, the better and faster the DD will work.
On the contrary, when the approximation is poor, the DD will not be advantageous with respect to the VG.
In fact, as $\overline{~\bf\chi ~}$ is positive definite, no matter how bad the approximation is,  for  sufficiently small $\alpha$ the DD will still converge toward the right solution ${\bf X^*}$.

In conclusion, for cases where the inference problem is meaningful and the consequently the approximation (\ref{chiApprox}) is valid, the DD algorithm will speed up the inference, whereas in the other cases DD will be useless but not counterproductive.


\section{The learning dynamics}
\label{sect::LearningDynamics}
All the  approaches explained before, (VG, NM and DD) solve the learning task through a discrete-time dynamics in the fields space. 
At each time step the estimation of the model averages of the observables (${\bf Q[X]}$) allows to compute the fields update and continue the dynamics.
For large system size $N$, however, the exponential complexity of the problem prevents the exact computation of the observables averages and some approximations are required.
A standard, but fruitful, choice consists in using Markov-Chain Monte-Carlo (MC) with Metropolis algorithm to sample $M$ system configurations and use them to estimate:
\begin{equation}
{\bf Q}_{\bf X}^\text{MC} \equiv \frac{1}{M} \sum_{b=1}^M {\bf \Sigma}(\sigma^\text{MC}(b)~)~. \label{MCaverage}
\end{equation}
${\bf Q}_{\bf X}^\text{MC}$ is now a random variable approximating ${\bf Q[X]}$ up to $O(\sqrt{1/M})$ fluctuations.

Even for large $M$, the randomness of ${\bf Q}_{\bf X}^\text{MC}$ will eventually affect the  convergence of the dynamics, as for ${\bf X}_t$ sufficiently close to ${\bf X^* }$, the size of the gradient will become comparable with its fluctuations.
However, at the beginning,  we expect the dynamics to be almost deterministic and then to become stochastic only at the end.
We indeed separate the dynamics in two regimes:
\begin{enumerate}
\item \textit{Approaching the convergence}, when ${\bf \overline{P} - {\bf Q}_{\bf X}^\text{MC} }$ is small but still much larger than its fluctuations.
\item \textit{The long-time stochastic dynamics}, when  $ {\bf \overline{P} - {\bf Q}_{\bf X}^\text{MC} }  \simeq  \sqrt{1/M}$ and thus comparable with its fluctuations.
\end{enumerate}
In appendices \ref{app::Approaching} and \ref{app::StochasticDynamics} we will provide several details of the two regimes and here we only present the mayor results.

\subsection{Approaching the convergence}
\label{sect::Approaching}

After an initial transient where the dynamics strongly depends on the chosen  algorithm (VG,NM or DD) and on the initial conditions, we expect the ${\bf X}_t$ to approach the log-likelihood maximum ${\bf X^* }$, so that we can approximate the log-likelihood function up to the quadratic order. 
Given $\delta  l[{\bf X}] \equiv l[{\bf X}] -  l[{\bf X^* }] $, we have
\begin{eqnarray}
\delta l[{\bf X}] &\simeq&  -\frac{B}{2}~  ({\bf X - X^*})~\cdot ~ \chi[{\bf X^*}] ~ \cdot ~ ({\bf X - X^*}) \nonumber  \\
&\approx& -\frac{B}{2} ~ ({\bf X - X^*})~\cdot ~ {\overline{{~ \chi}~}  }  ~ \cdot ~ ({\bf X - X^*})    ~.
\label{logLlongTime}
\end{eqnarray}  
In this approximation the dynamics is exactly solvable upon projecting the fields on the $\chi[{\bf X^*}]$ Eigenvectors $\{ {\bf{V}^\mu}\}_{\mu=1}^D$: $\d X^\mu_t = \sum_a V^\mu_a \d X_{a,t}  $. 

Along a $\mu$-Eigenspace the convergence of the VG algorithm scales with the corresponding Eigenvalue $\l_\mu$ as: $\d X^\mu_t \sim (1- \alpha \l_\mu)^t$.
Consequently, to ensure the algorithm convergence,  we need $\a < 2 / \l_\mu$ along all directions.
Moreover, by optimizing the convergence speed along all directions simultaneously we obtain $\a^{\text{(VG)}}_{BEST} = 2/( \l_>+\l_<)$, where $\l_{>/<}$ are the largest/smallest Eigenvalue.
In App. \ref{app::Approaching} we present some details and here we simply notice how $\a^{\text{(VG)}}_{BEST} $ can be squeezed to very small values by a large $\l_>$ preventing the learning along all the direction with $\lambda_\mu \ll \l_>$.
Consequently, for dataset where the Eigenvalues of $\chi[{\bf X^*}]$, or of its approximation $\overline{{~ \chi}~}$, spread over several order of magnitude the convergence of the VG algorithm will be very slow. 

Because of the quadratic approximation, the NM and the DD algorithms coincide and  we do not distinguish between them.
Within their dynamics, $\d X^\mu_t \sim (1- \alpha)^t$ independently of the Eigenvalues $\l_\mu$.
consequently $\a <2$ is the only convergence condition and $\a^{\text{(DD)}}_{BEST} = 1$.




\subsection{The long-time stochastic dynamics}
\label{sect::StochasticDynamics}

Before proceeding we like to introduce the shortcut notation:
\begin{equation}
\mathcal{N} [{\bf m ;s}](t) \equiv \frac{ e^{ - \frac{1}{2} \sum_{ab} (t_a-m_a) s_{ab}^{-1}  ( t_b-m_b)  } }{  \sqrt{ (2 \pi)^D | {\bf s}| ~} }~,
\end{equation}
a normal distribution with average ${\bf m}$ and covariance ${\bf s}$ evaluated at $t$.

When ${\bf X \to X^*}$, we expect ${\bf \nabla }  l_{\bf X}^\text{MC} \equiv B \big( \overline{\bf P} - {\bf Q}_{\bf X}^\text{MC} \big) \to 0$ only on average, with not negligible fluctuations of $O(\sqrt{1/M})$ .
In this regime the fields dynamics is not anymore deterministic and the convergence becomes a stochastic process.
Consequently, rather than converge to a fixed point, the fields will approach an equilibrium stationary regime around ${\bf X^* }$. 

The stochastic process is ruled by the discrete-time master equation:
\begin{equation}
P_{t+1}({\bf X'})  =  \int D{\bf X}~ P_t({\bf X}) ~W_{{\bf X}\to {\bf X'}} \label{masterEq},
\end{equation}
where the transition rates depend on the distribution of ${\bf \nabla }  l_{\bf X}^\text{MC}$. 

In the large $M$ limit, the gradient distribution can be approximated as a Gaussian  to obtain an analytic expression of $W_{{\bf X}\to {\bf X'}}$, see app. \ref{app::StochasticDynamics}.
By asking $P_t({\bf X})$ to be invariant under the evolution (\ref{masterEq}) we can obtain the stationary distribution $P_\infty({\bf X})$:
\begin{eqnarray}
P^\text{VG}_\infty( {\bf X} )   &=& \mathcal{N} \Big[ {\bf X^*} ; \frac{\a}{M } \big(2 {\bf \delta_D}- \alpha \overline{ \bf ~\chi ~ } \big)^{-1} \Big]( {\bf X} ), \label{PSofXVG} \\
P^\text{DD}_\infty( {\bf X} )   &=&  \mathcal{N} \Big[ {\bf X^*} ; \frac{\a}{M (2 - \alpha)} \overline{\bf ~\chi ~ }^{-1} \Big]( {\bf X} ), \label{PSofXDD}
\end{eqnarray}
where $\delta_D$ is the identity matrix in dimension $D$. Here the typical fluctuations of ${\bf X}$ around ${\bf X^*}$ must consistently verify the approximation ${\bf X \approx X^*}$: 
$\la {\bf (X - X^*)^2} \ra_{P_\infty }$ should be small enough to allow the expansion (\ref{logLlongTime}).

In the stationary regime, on top of the fluctuations induced by the MC, the ${\bf X}$ distribution will induce a second source of noise in the actual distribution of ${\bf Q}^\text{MC}$.
On average we expect (see app. \ref{app::StochasticDynamics}):
\begin{eqnarray}
P^\text{VG}_\infty( {\bf Q}^\text{MC} ) &=& \mathcal{N} \Big[  \overline{\bf P}; \frac{2 \overline{\bf ~\chi~} \left(2 \d_D - \a \overline{ \bf ~\chi ~ } \right)^{-1}}{M} \Big]( {\bf Q}^\text{MC} ) \label{PVGSofGrad} \\
P^\text{DD}_\infty( {\bf Q}^\text{MC} ) &=& \mathcal{N} \Big[  \overline{\bf P}; \frac{2  \overline{ \bf ~\chi ~ } }{M (2-\a)}\Big]( {\bf Q}^\text{MC} ) \label{PDDSofGrad}~.
\end{eqnarray}
Interestingly, the two algorithms provide different ${\bf Q}^\text{MC}$ distributions.
In particular, the VG fluctuations along directions with large (small) $\l_\mu$ are larger (smaller) than the DD ones.
These differences will have consequences on the ability of both algorithms to reproduce the dataset statistics and thus to avoid both under- and  over-fitting. 

\subsection{The dynamics under an external stochastic force}
\label{stochForce}

For forthcoming purposes, here we characterize the learning dynamics when a linear stochastic force term is added to the gradient term in the learning rules.
This analysis will be useful when a $L2$ prior term is included in the inference procedure.
We modify eq.  (\ref{ApproximatedNewton}) as:
\begin{equation}
 {\bf \d X }^{\text{DD}_\eta}_t = \alpha~ {\overline{{~\bf \chi_\eta}}  }^{-1} \cdot  \big(~ {\bf \nabla } l_{\bf X}^\text{MC} + F^\eta_{\bf X} ~\big)~. 
\end{equation}
where
\begin{eqnarray}
\overline{{~\bf \chi_\eta}} &\equiv&  \overline{{~\bf \chi}~} + {\bf \eta} \label{ChiEta}\\
P\big(F^\eta_{\bf X}\big) &=&   \mathcal{N} \Big[ -\eta \cdot X ; \frac{ {\bf \eta}  }{M}\Big]( F^\eta_{\bf X} ). \label{pOfFx}
\end{eqnarray}
The calculation for the stationary fields distribution follows as in the previous section and it results in:
\begin{equation}
P^{\text{DD}_\eta}_\infty( {\bf X} )   =  \mathcal{N} \Big[ {\bf X_\eta^*} ; \frac{\a}{M (2 - \alpha)} \overline{\bf ~\chi_\eta  }^{-1} \Big]( {\bf X} ), \label{PSofXDDeta}
\end{equation}
where
\begin{equation}
{\bf X_\eta^*} =   \overline{\bf ~\chi_\eta  }^{-1} \cdot \overline{{~\bf \chi}~} \cdot  {\bf X^*}. \label{xEta}
\end{equation}
Analogously to (\ref{PDDSofGrad}),  from (\ref{PSofXDDeta}) it follows:
\begin{equation}
P^{\text{DD}_\eta}_\infty( {\bf Q}^\text{MC} -F^\eta ) =  \mathcal{N} \Big[  \overline{\bf P} ; \frac{2  \overline{ \bf ~\chi_\eta } }{M (2-\a)}  \Big]( {\bf Q}^\text{MC}-F^\eta ) ~.\label{PDDSofGradeta}
\end{equation}
Expressions for the VG algorithm can be easily obtained from (\ref{PSofXDDeta})  and (\ref{PDDSofGradeta}) with substitution $\alpha \to \alpha \overline{ \bf ~\chi_\eta }^{-1}$.

\section{Avoiding under- and over-fitting by sampling from the posterior distribution.}
\label{avoidingOverfitting}

In the previous section we shown how any actual implementation of the algorithm is stochastic.
In particular, depending on the chosen algorithm and its parameters, $\alpha$ and $M$, we expect a whole probability distribution on the output fields ${\bf X}$.
This stochasticity raises questions on which implementation should be preferred in practical applications.
The simple strategy of trying to obtain the best possible approximation of ${\bf X^*}$ by increasing $M$ or reducing $\alpha$ will unambiguously lead to data over-fitting thus limiting the model prediction power.
The presence of noise in the estimation of $\overline{\bf P}$ induces some uncertainty on the model fields ${\bf X}$ that should be taken into account to avoid over-fitting.

This fields uncertainty is quantified by the posterior distribution of the fields given the data, see eq. (\ref{posterior}). 
Within the approximation (\ref{logLlongTime}), the posterior simplifies to
\begin{equation}
P^\text{Post}\big(~ {\bf X} ~\big| ~\text{data} ~ \big) \propto e^{ -\frac{B}{2} ({\bf X - X^*})\cdot {\overline{{~ \chi_{ab}}~}  } \cdot ({\bf X - X^*})  } \label{PostFlat}
\end{equation}
in the case of flat prior and to
\begin{eqnarray}
P^\text{Post}\big(~ {\bf X} ~\big| ~\text{data} ~ \big) &\propto& e^{  -\frac{B}{2}\left[~ ({\bf X - X^*})\cdot {\overline{{~ \chi}~}  } \cdot ({\bf X - X^*}) + {\bf X} \cdot {\bf \eta}  \cdot {\bf X}~\right] } \nonumber \\
&\propto&  e^{  -\frac{B}{2} ({\bf X - X_\eta^*})\cdot {\overline{{~ \chi_\eta}}  } \cdot ({\bf X - X_\eta^*}) } \label{PostL2}
\end{eqnarray}
when an $L2$ prior, see eq (\ref{L2prior}), is considered (${\bf X_\eta}$ and $\overline{~ \chi_\eta}$ are those defined in eqs. (\ref{xEta}) and (\ref{ChiEta})). 

If by tuning the algorithm implementation and settings we can match the a stationary fields distribution and the posterior, we will be able to sample from it, thus avoiding any over- and/or under-fitting.
In case of flat prior, we have to compare (\ref{PostFlat}) with (\ref{PSofXVG}) and (\ref{PSofXDD}), whereas in case of $L2$ prior (\ref{PostL2}) with (\ref{PSofXDDeta}) for the DD algorithm and the analogous for the VG one: by setting
\begin{equation}
\alpha = \frac{2 M}{B+M}
\label{alphaOpt}
\end{equation}
the DD algorithm will sample from the correct posterior distribution and over- and under-fitting will be avoided along all the $D$ dimensions.
On the contrary, for the VG algorithm  such a setting does not exist and the algorithm will not sample from the posterior.
In particular directions with $\a \l_\mu \ll 2 M/(B+M)$ are strongly over-fitted and those with $\a \l_\mu \gg 2 M/(B+M)$ are strongly under-fitted.
Otherwise stated: by choosing the VG algorithm with $\alpha$ small enough to well reproduce directions with large $\lambda_\mu$ will result in overfitting of all directions with small $\lambda_\mu$.

Among the consequences of overfitting, for finite $B$ we expect an overestimation of the observed model log-likelihood.
In fact, as the exact inference fields ${\bf X^*}$ reproduce also the noise in the experimental averages $\overline{\bf P}$, we expect $l^* \equiv l[{\bf X^* }] > \wh{l}$, where $\wh{l}$ is the log-likelihood of the model that generates the data, the true log-likelihood.
In order to quantify this effect, in appendix \ref{logL_comparison} we compute the average log-likelihood estimation in the case of the exact inference when the data are synthetically generated by a MaxEnt model with true fields $\wh{\bf X}$.
By averaging over the distribution of $\overline{\bf P}$, we find that
\begin{equation}
\langle l^* \rangle - \wh{l} = \frac{D}{2B}~,
\end{equation}
which shows how the log-likelihood maximization, see eq. (\ref{maximization}), induces a finite bias leading to a log-likelihood overestimation.
In the case of the DD algorithm, instead, the average over the posterior distribution exactly cancels the bias, and the true log-likelihood value is recovered.

\section{Algorithm implementation}
\label{set::algorithm}

The core of DD algorithm is to iteratively update the fields ${\bf X}_t$ with the rule (\ref{ApproximatedNewton}), where ${\bf Q[X_t]}$ is approximated by  $ {\bf Q}_{\bf X}^\text{MC}$ through  a MC sampling of $M$ configurations.
Ideally $\alpha = \alpha^\text{DD}_\text{BEST} = 1$ and $M=B$ is the fastest setting that satisfies the condition (\ref{alphaOpt}).
However any practical implementation will face two main difficulties.
The first lies in the fact that the quadratic approximation (\ref{logLlongTime}) may not be valid.
This will happens at the beginning of the dynamics, when  ${\bf X}_t$ if far from  ${\bf X^*}$, but also when $B$ is not large enough to have $({\bf X}_t - {\bf X^*})^2$ small enough to discard third order terms.
If it is the case, the distribution (\ref{PSofXDD}) will be non-stationary.
For this reason we allow the algorithm to adapt the value of $\alpha$ at each time-step.
The second difficulty reflects the fact that the algorithm  is not suppose  to converge to  ${\bf X^*}$ but rather to a probability distribution. 
Consequently we need a condition that signals the onset of the thermalization and allows us to start storing ${\bf X}_t$s as samples from the posterior distribution.
For this reason we introduce the following quantity:
\begin{equation}
\epsilon_t \equiv \sqrt{ \frac{B}{2 D}  \Big( {\bf ~\overline{P}- Q_{\bf X_t}^\text{MC} } \Big) \cdot \overline{\bf ~\chi~}^{-1} \cdot  \Big({\bf ~\overline{P}-Q_{\bf X_t}^\text{MC} }\Big)} ~. \label{epsDef}
\end{equation}
Under the distribution (\ref{PDDSofGrad}) with $\alpha = 1$ and $M=B$, $\epsilon_{\infty}$ will have distribution:
\begin{equation}
  P(\e_{\infty}) = \frac{2~ D^\frac{D}{2}}{\G(\frac{D}{2})~ 2^\frac{D}{2}} \e_{\infty}^{D-1} ~\exp\Big\{-\frac{D}{2} \e_{\infty}^2\Big\}
\end{equation}
where $\G(x)$ is the Gamma function.
$\epsilon_\infty$  has:
\begin{eqnarray}
\la \e_{\infty} \ra &=& \sqrt{ \frac{2}{D} }~ \frac{\G(\frac{D+1}{2})}{\G(\frac{D}{2})} \xrightarrow[D \to \infty]{} 1 \\
\sqrt{\la \e_{\infty}^2\ra - \la \e_{\infty} \ra^2 } &=& \sqrt{1 -\la \e_{\infty} \ra^2~ }~ \xrightarrow[D \to \infty]{} 0~.
\end{eqnarray}
In the large $D$ limit, the $\epsilon_{\infty}$-distribution shrinks to a Dirac-delta  function at $\epsilon_{\infty}=1$:
if before thermalization we expect with high probability $\e_t \gg 1$, once the algorithm starts sampling from the stationary distribution (\ref{PSofXDD}) we expect $\e_t\approx 1$.
Consequently, once the condition $\e_t \leq 1$ is full filled, the subsequent ${\bf X}_t$ are good estimations of the fields.
In order to effectively sample from  (\ref{PSofXDD}) it will be still necessary to keep running the algorithm in order to decorrelate from the initial condition ${\bf X}_0$.

The DD algorithm that we implemented can be sketched as follows:
\begin{enumerate}
\item Chose an initial configuration for ${\bf X_0}$ compute/evaluate ${\bf Q[X_0]}$, see eq. (\ref{modelAverage}) and compute $\e_0$ from eq. (\ref{epsDef}). Then set  $\alpha_0=1$ and  $M_0 = \min(\frac{B}{\e_0^2 },B)$. 
\item Iterate the following step:
\begin{enumerate}
 \item update the ${\bf X_t}$ through eq. (\ref{ApproximatedNewton}), 
 \item estimate ${\bf Q[X_t]}$ though $M_t = \min( \frac{B}{\e_{t-1}^2 } ,B)$ Gibbs samplings, see eq. (\ref{MCaverage}),
\item compute $\e_t$ from eq. (\ref{epsDef}) and
\begin{enumerate}
\item  if $\e_t < \e_{t-1}$  accept the update and set $\a_{t}=\a_{t-1} \d^+$
\item  otherwise  discard the update, set $\a_{t}=\alpha_{t-1} / \d^-$ and estimate again ${\bf Q[X_t]}$.
\end{enumerate}
\end{enumerate}
\item As soon as the condition $\epsilon_t <1$ is full filled,
\begin{enumerate}
\item either fix $\alpha$ and let the distribution $P_t({\bf X})$ decorrelate from ${\bf X}_0$ and thermalize to $P_\infty({\bf X})$, see (\ref{PSofXDD}).
\item either stop the algorithm and retain ${\bf X}_t$ as a fields list solving the inference problem.
\end{enumerate}
\end{enumerate}  
A variable and adapting $\a$ is required because the system can be be far outside  the validity range of (\ref{logLlongTime}).
In order to avoid cycles, we heuristically set  $\d^+=1.05$ and $\d^-=\sqrt{2}$.

The choice of an adapting $M_t=\min(\frac{B}{\e_t^2 },B) \leq B$, instead,  allows to save time during the deterministic dynamics regime, see sect. \ref{sect::Approaching},  when the algorithm does not require a high precision in the gradient estimate. 
$\e_t$, in fact measures also the norm of the log-likelihood gradient in the metric defined by its fluctuations, $B^2 I[{\bf X}]/M = B^2 \chi[{\bf X}]/M\approx B^2 \overline{~\bf \chi~}/M$, see eq. (\ref{pGradGivenX}): 
\begin{equation}
\e_t = \sqrt{\frac{M}{B}} \norm{ {\bf \nabla}  l[{\bf X_t}]}_{B^2 \overline{~\bf \chi~}/M }~.
\end{equation}
For  $M_t=B/\e_t^2$ we expect the estimate of the gradient to be statistically confident within one standard deviation and consequently enough well estimated.
Moreover the randomness induced by small $M_t$ values decorrelates the dynamics from the initial condition.  
When the algorithm approaches convergence, $\e_t \to 1$ and consequently $M_t \to B$ thus  full filling condition (\ref{alphaOpt}).

The complexity of the algorithm depends linearly on $N B$ through the MC sampling.
From the tests we performed the number of required steps depends mostly on the dataset properties and in particular on the exactness of ${\bf \chi}[{\bf X^* }] \approx \overline{~\bf\chi ~}$.  
The hardest limit of the DD lies in the memory allocation for storing $\overline{ ~ \bf\chi ~ }$.
Working in double precision, for an hardware with $32Gb$ of RAM the maximum number of manageable units is $N \sim 350$\footnote{As the matrix $\overline{ ~ \bf\chi ~ }$ has several symmetries a compressed encoding could decrease the required storage memory}.

\subsection{The $L2$ prior term}
In order to include a $L2$ prior term we add to the dynamics the stochastic force term introduced in sect. \ref{stochForce}.
At each iteration, we should modify the update rule by adding to the gradient term ${\bf \nabla } l_{\bf X}^\text{MC}$ an independent realization of the random variable ${\bf F^\eta_X}$ according to the distribution (\ref{pOfFx}).
As the prior term will biases the inference we need to modify the function (\ref{epsDef}) in order to account for the difference between (\ref{PDDSofGrad}) and (\ref{PDDSofGradeta}):
\begin{equation}
\epsilon_t^\eta = \sqrt{ \frac{B}{2D} \big( {\bf P- Q +F^\eta_{\bf X_t}}  \big)\cdot \overline{~\chi_\eta}^{-1} \cdot \big({ \bf P-Q + F^\eta_{\bf X_t}} \big)}
\end{equation}
which, remarkably, does not require the explicit inversion of the matrix $\overline{\bf ~\chi~}$. 
Moreover, as $\overline{\bf ~\chi~}$ is, by construction, non-negative, $\overline{\bf ~\chi_\eta}$ is a positive matrix and can be inverted.
As a consequence, the DD algorithm regularized with $L2$ prior can be applied when an undersampling induces zero modes in the empirical estimation of the susceptibility matrix.

\section{Test}
\label{sect::Test}
\begin{figure}[t]
\includegraphics[clip=true,keepaspectratio,angle=-0,width=1.0\columnwidth]{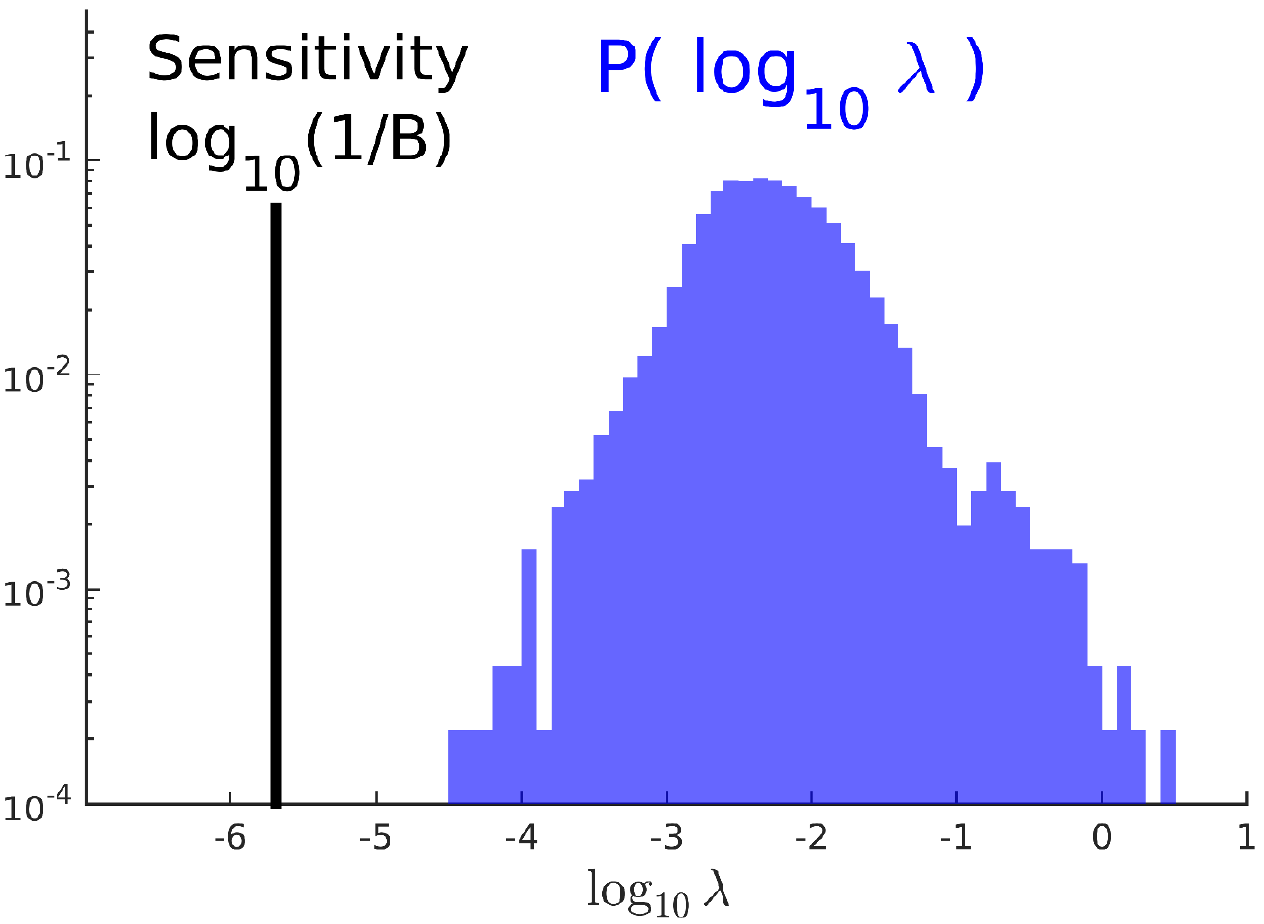}
\caption{(Colors online) Eigenvalues distribution of the empirical susceptibility matrix, see eq. (\ref{chiApprox}), for a pairwise Ising model inference problem on biological data.
As often happens in biological problems \cite{Machta13}, the Eigenvalues span several order of magnitude enhancing the heterogeneity of the parameter space.
As will be show in the inset of Fig. \ref{fig:correlations}, the susceptibility matrix of the inferred Ising model has a very similar spectrum.
Consequently, even with the optimal learning rate, $\a^{\text{(VG)}}_{BEST}$ of eq. (\ref{alphaBest}), VG will be very slow in learning  along the directions with the smallest Eigenvalues.
As the sensitivity ($1/B$) lies well below the smallest Eigenvalue all the susceptibility spectrum is informative and  has to be reproduced by the inference. See text for more details.       
Data from a $\sim \!2.1h$ (binned at $16\text{ms}$) recording of $95$ rat retinal ganglion cells subject to visual stimulation \cite{Mora15}.
}
\label{fig:fisherEv}
\end{figure}

\begin{figure}[t]
\includegraphics[clip=true,keepaspectratio,angle=-0,width=1.0\columnwidth]{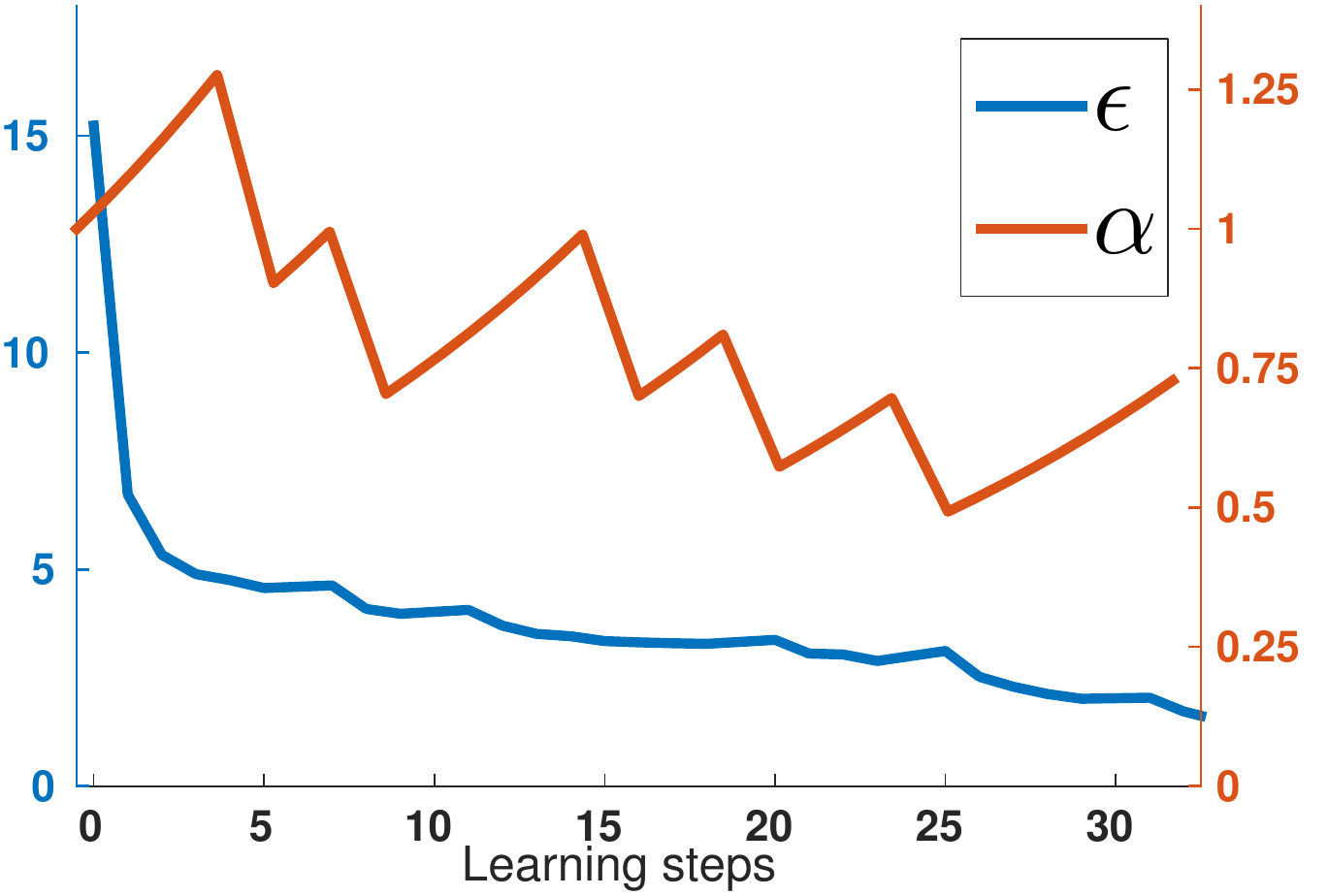}
\caption{(Colors online) Behavior of the pairwise Ising model inference of biological data through the Data-Driven algorithm.
$\e_t$ (lower blue curve, left axis) and of $\alpha_t$ (upper red curve,right axis) during the algorithm running plotted against the iterative step number.
Abscissa is  not proportional to the running time as each step performs a different number $M$ of MC sampling, see text.
$\a_t$ is raised (lowered) at every step where $\e_t$ decreases (increases).
The algorithm requires $100 \pm 5 s$ to produce a set of fields ${\bf X}$ satisfying the early-stop condition $\e_t<1$.
Dataset as in fig. \ref{fig:fisherEv}.
}
\label{fig:behavior}
\end{figure}

\begin{figure}[t]
\includegraphics[clip=true,keepaspectratio,angle=-0,width=\columnwidth]{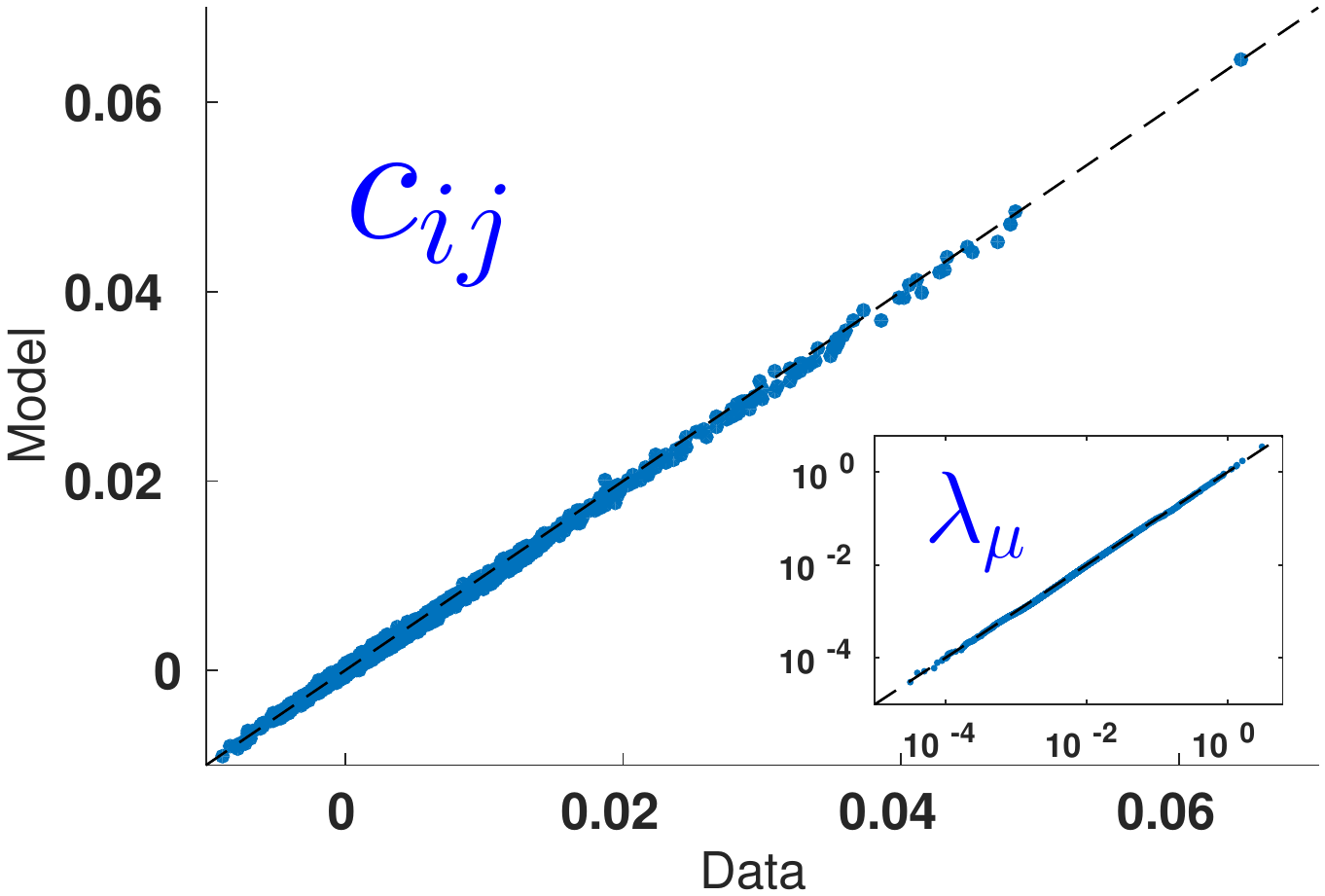}
\caption{(Colors online) Scatter-plot of the experimental connected correlations, $c_{ij}=\langle \s_i \s_j \rangle_\text{DATA}-\langle \s_i \rangle_\text{DATA} \langle \s_j\rangle_\text{DATA}$, against those estimated with MC sampling of $M =B = 4.8~10^5$ equilibrium configurations of the inferred model distribution.
Inset: scatter-plot of the ordered Eigenvalues of $\overline{\bf ~\chi~}$ (x-axis) against those of ${\bf \chi[X^*]}$.
Dataset as in fig. \ref{fig:fisherEv}.
}
\label{fig:correlations}
\end{figure}

\begin{figure}[t]
\includegraphics[clip=true,keepaspectratio,angle=-0,width=0.99\columnwidth]{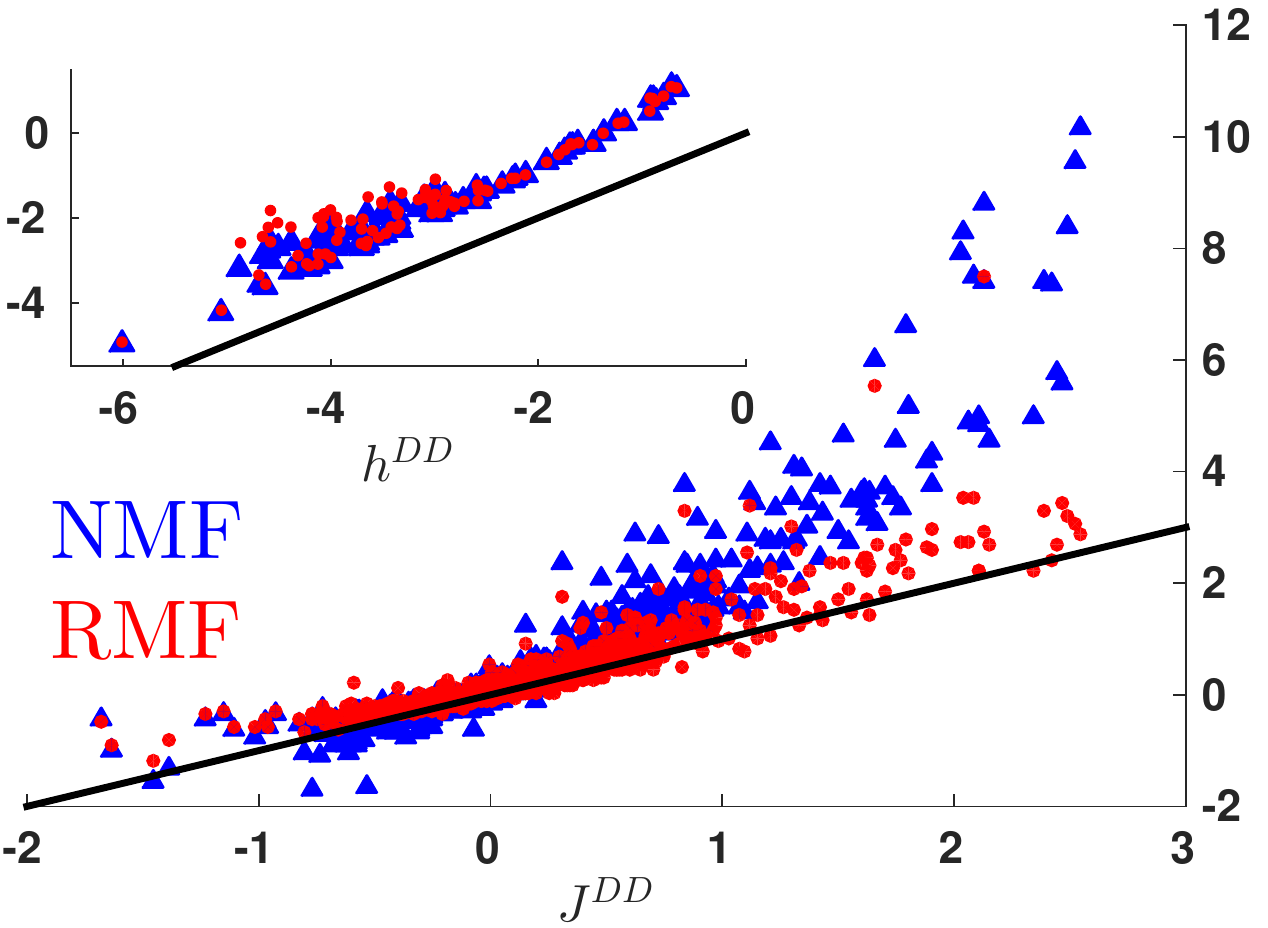}
\caption{(Colors online) Scatterplots of the inferred couplings (main panel) and biases (inset) with the DD algorithm against those of Naive (blue triangles) and Resummed (red circles) Mean-Field.
The black lines show equality.
All the approximations fails in the  reconstruction with a tendency to overestimate both couplings and fields (notice the different axis scale).
Dataset as in fig. \ref{fig:fisherEv}.
}
\label{fig:couplings}
\end{figure}

\begin{figure}[t]
\includegraphics[clip=true,keepaspectratio,angle=-0,width=\columnwidth]{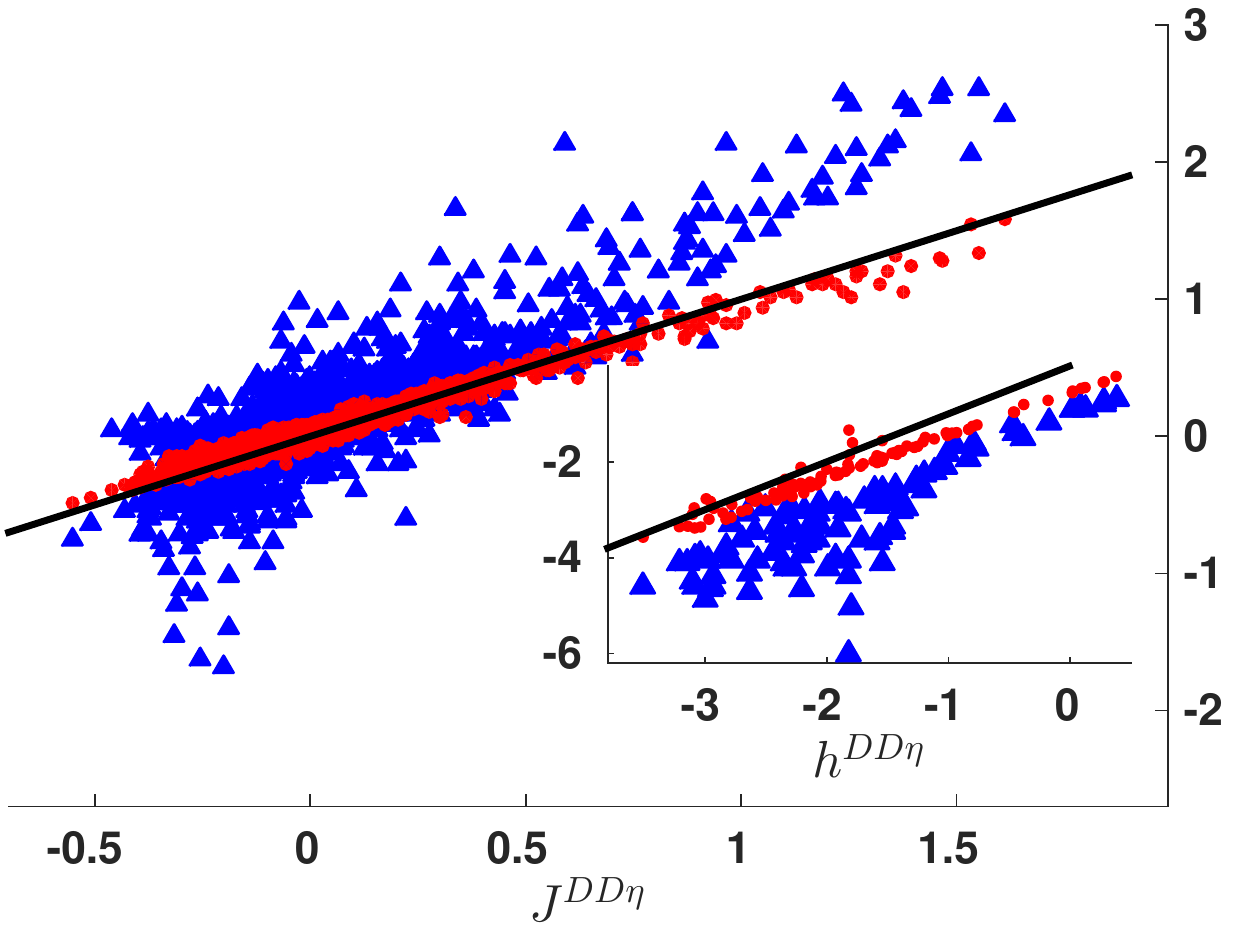}
\caption{(Colors online) Scatterplot of the inferred couplings (main panel) and biases (inset) for a pairwise Ising model with and without  $L2$ prior term, see eq. (\ref{L2prior}), with ${\bf \eta} = 5 \cdot 10^{-3}~ {\bf \delta_D}$.
Blue triangles compare the inferred fields  with (x-axis) and without (y-axis) prior term. Red circles compare the inferred fields with regularization (x-axis) with the expected ones (y-axis):
${\bf X_\eta^{DD}} \equiv \overline{~\chi_\eta}^{-1} \cdot \overline{~\chi ~} \cdot {\bf X^{DD}}$, see eq. (\ref{xEta}). 
The black line shows equality.
As a consequence of the $L2$ regularization the inferred fields are smaller in absolute value.  
Dataset as in fig. \ref{fig:fisherEv}.
}
\label{fig:regScatter}
\end{figure}

As explained before, the DD algorithm performance depends mostly on how good is the chosen MaxEnt model in modeling the dataset statistical properties.
As expected, we succeed in inferring back the fields ${\bf X}$ from synthetic dataset obtained through simulation of MaxEnt models.
Indeed we find more interesting to present an application to biological data where the underlying statistical distribution does not belong to class of models considered in the Inference task. 
 
We tested the DD algorithm on an \textit{ex-vivo} multi-electrode array recording \cite{Marre12} of 95 rat retinal ganglion cells \cite{Mora15}.
The retina was stimulated though a video showing two randomly moving bars \cite{Mora15} displayed with a frame rate of $60Hz$. 
The $\sim \!2.1h$ spike trains recording was binned at $16\text{ms}$ to obtain $B \sim 4.8 \cdot 10^5$ system configurations, where we assign $\sigma_i(b)=1$ if cell $i$ emitted at least one action potential in the time-bin $b$ and $\sigma_i(b)=0$ otherwise.
We infer a pairwise Ising model thus restricting the observables list to single and pairwise correlations ($\{ \Sigma_a\}_{a=1}^D = \{ \{\s_i\}_{i=1}^N ,  \{\s_i \s_j\}_{i<j=1}^N\}$) and consistently the fields to biases and pairwise interactions ($\{ X_a\}_{a=1}^D = \{ \{h_i\}_{i=1}^N ,  \{J_{ij}\}_{i<j=1}^N\}$).
With this observables choice the model takes the form of the well known Disordered Ising Model:
\begin{equation}
P_{\bf h,J}( {\bf \s}) = \exp \Big\{ \sum_i h_i \s_i + \sum_{i<j} J_{ij} \s_i \s_j \Big\}~  / ~ Z[ {\bf h},  {\bf J} ]
\end{equation} 
where $Z[ {\bf h},  {\bf J} ]$ is the normalization constant.

Before performing the inference it is important to check whether the the number of measurements is large enough to empirically estimate the observables averages with good precision.
In the large $B$ limit, we expect an empirical error ${\bf \delta P}$ in the estimate of $\overline{\bf P}$ to have zero mean and covariance $\hat{~\bf \chi~}/B$, see eq. (\ref{pOfDeltaP}).
By linear regression, en error ${\bf \delta P}$ in the data will induce en error:
\begin{equation}
{\bf \delta X} = \left( \left. \frac{\bf \delta Q[X]}{\bf \delta X}\right|_{\bf X=X^*} \right)^{-1} \cdot {\bf \delta P} \approx \hat{~\bf \chi~}^{-1} \cdot {\bf \delta P}
\end{equation}
on the inferred fields.
Consequently, by approximating $\hat{~\bf \chi~} \approx \overline{~\bf \chi~}$:
\begin{equation}
P({\bf \delta X}) = \mathcal{N}\big[~ 0 ~; ~ (B \overline{~\bf \chi~})^{-1} ~]({\bf \delta X})~.
\end{equation}
To ask for this error to be small, we should indeed require:
\begin{equation}
\sqrt{ B~\lambda_{\mu}  } \gg 1~,\quad \text{for all }\mu~ \Rightarrow ~ \frac{1}{B} \ll \lambda_\text{MIN}~.
\label{sensitivityCondition}
\end{equation}
As shown in Fig. \ref{fig:fisherEv}, in the considered dataset, condition (\ref{sensitivityCondition}) is satisfied: the experimental sensitivity, $1/B$ is much smaller then the smaller Eigenvalue of $\overline{~\bf \chi~}$.

On an $\text{Intel}^\circledR  ~ \text{Core}^\text{TM}~ \text{i7-4770}$ with eight cores at $3.4 \text{GHz}$ the written in $\text{Matlab}^\circledR$ DD algorithm  takes $100 \pm 5 s$ to reach convergence.
In Fig. \ref{fig:behavior} we  show the behavior of both $\epsilon_t$ and $\alpha_t$ against the number of inference steps. 
As the number $M$ of MC samplings is not constant during the inference, steps are of different time duration.

In order to show the quality of the inferred fields, in the main panel of Fig. \ref{fig:correlations}  we scatter-plot the experimental connected correlations against those estimated through $B$ sampling of the inferred Ising model distribution.
In order to give an insight on the validity of (\ref{chiApprox}), in the inset we scatter-plot the ordered Eigenvalues of $\overline{\bf ~\chi~}$ (x-axis) against those of ${\bf \chi[X^*]}$.

To convince the reader that the tested inference problem was difficult, we apply two example Mean-Field approaches to the dataset.
In Fig. \ref{fig:couplings} we show the scatterplot of the inferred fields against those of Naive \cite{Kappen97} and Resummed \cite{Jacquin15} mean-field approximations.
Despite Resummed works much better than Naive, for this dataset both approximations largely overestimate couplings and biases.

For comparison, we tested the VG algorithm on the same dataset with $M=B$ and various $\a \leq \a^{\text{(VG)}}_{BEST}$.
With $\a = 0.2\a^{\text{(VG)}}_{BEST} $ the VG takes $\sim 4.2 \cdot 10^4 s$ to converge, whereas for larger $\a$ values it was not able to satisfies the early-stopping condition $\epsilon<1$. 

In Fig. \ref{fig:regScatter} we scatter plot in blue the inferred fields against those inferred with a $L2$ regularization term, see eq. (\ref{L2prior}), with ${\bf \eta} = 5 \cdot 10^{-3}~ {\bf \delta_D}$, 
whereas in red we scatter the inferred  with prior fields against the expected ones, see eq. (\ref{xEta}).

A part from the retina example, we succeeded in inferring the Ising model fields from synthetic data, human temporal cortex recording \cite{Peyrache12} , rat pre-frontal cortex recording \cite{Peyrache09} and others. 
In all the tested cases convergence times are of the order of tens to few hundreds of seconds.

\section{Conclusions and discussion}

In this study we introduced a Markov-Chain Monte-Carlo based algorithm for solving the MaxEnt inference problem and sampling from the posterior probability distribution of the MaxEnt fields ${\bf X}$, namely couplings and biases of an Ising model or their generalization in the presence of non-pairwise interactions.
We carefully analyze the learning dynamics and separate two different regimes: i) a deterministic dynamics that approaches the solution and ii) a stochastic dynamics of the probability distribution of the fields, $P_t({\bf X })$, that thermalizes to a stationary distribution.
By tuning the algorithm settings, namely $\a$ and $M$, this distribution can reproduce the posterior distribution of the inference fields thus allowing to sample.
We concluded presenting an implementation of the algorithm and a test on a biological dataset showing how the presented algorithm outperforms the standard learning approach.
The core of the algorithm is the approximation (\ref{chiApprox}) which requires at first to have enough data to properly estimate the empirical susceptibility then that the probability model chosen for the inference reproduces quite well the data statistics. 
As the applications to biological data have shown, both conditions have not to be intended as rigid constraints, but rather as requirements for obtaining the largest advantage of the DD approach.

The key properties of the MaxEnt inference that driven this study are the relationships (\ref{equalities}).
The equality among the log-likelihood Hessian and the model Fisher allows to tune the MC fluctuations of the log-likelihood gradient (ruled by the Fisher) in order to reproduce the fluctuations of Posterior distribution around the log-likelihood maximum (ruled by the Hessian).
The equality with the susceptibility matrix allows to obtain an expression of both the Fisher and Hessian that depends only on ${\bf X}$-dependent averages of ${\bf X}$-independent functions that at the solution ${\bf X^*}$ can be approximated through the data.

In sect. \ref{sect::LearningDynamics}, we have carefully analyzed  the learning dynamics and in \ref{sect::StochasticDynamics} we have characterized it as a stochastic process that converges to a stationary distribution.
Later, in sect. \ref{avoidingOverfitting},  we have shown how the DD algorithm stationary distribution can be tuned to the posterior distribution.
This characterization has several advantages: 
\begin{itemize}
\item the computation of $P^\text{DD}_\infty( {\bf Q}^\text{MC} )$, together with the running evaluation of $\epsilon_t $, see eq. (\ref{epsDef}) provides an useful early-stopping condition for the algorithm that allows to check when the inference has been accomplished. Consequently it is useful to save much computational time. 
\item any inference algorithm has to account for the randomness of the MCMC fluctuation and typically large computational efforts are required to get rid of this noise source (large number of MCMC samplings $M$) . 
The DD algorithm, instead, takes advantage of these fluctuations to avoid overfitting and to decorrelate from the initial condition.
\item as we have shown in appendix \ref{logL_comparison}, in comparison with the exact inference, the posterior sampling allows to avoid overfitting the model.
\item as in the case of the VG algorithm, an overfitting along several directions happens with all the algorithms that do not \textit{rectify} the fields  space.
\item as explained in sect. \ref{sect::Bayes}, the posterior sampling can provide several consistent lists of inferred fields, thus allowing to test the robustness against the dataset noise of any analysis based on the inferred fields. 
\end{itemize} 

Our approach and the use of the MC randomness to induce noise in the algorithm outcome may remind the application of stochastic gradient method for sampling from the fields posterior distribution \cite{Welling11}.
Despite the results may look similar, in order the induce fluctuations, here we take advantage from the intrinsic algorithm randomness instead of introducing it by sub-sampling the dataset.
The mayor advantage results from the possibility to select and optimize properly the algorithm setting in order to get rid of the MC induced noise that otherwise will affect the dynamics in a spurious way.
One important drawback arises from the fact that the algorithm fluctuations may not reproduce the experimental ones when the Gaussian approximation (\ref{logLlongTime}) is not valid and second order terms are not enough to approximate the log-likelihood function around the solution.
In this case, however, one can notice how the equalities (\ref{equalities}) can be extended up to higher order cumulants. 
These relationships, together with an appropriate update rule that takes into account higher order corrections may extend the equivalence between $P_\infty({\bf X})$ and the posterior.
We let this generalization for forthcoming investigations.

The DD approach takes its place in the list of algorithms for exactly solving the MaxEnt model inference problem.
Its greatest strengths are the velocity and the possibility to sample from the posterior, whereas its stronger limitation is the memory requirement for the storage of $\overline{~\chi~}$.
As an example, the inference of a $N=350$ pairwise Ising model requires $32Gb$ of RAM.

Depending on the dataset properties the DD should or should not be preferred to others algorithms as \cite{Broderik07}, \cite{Cocco11} or \cite{Sohl-Dickstein11}.
In general, if the condition (\ref{sensitivityCondition}) is largely satisfied, we expect DD to be the best choice, because the approximation (\ref{chiApprox}) is expected to be valid.
In the opposite case, a largely unsatisfied (\ref{sensitivityCondition}) suggests that the inference of the chosen MaxEnt model is not meaningful and the observables list has to be modified.  
However for cases in between some tests with different algorithms have to be performed.
In particular cases, some available \textit{a priori} knowledge of the system could help in the algorithm  choice.
For example, if the underlying interaction graph is naturally clusterized in almost non-interacting subcomponents, Selective Cluster Expansion (SCE) \cite{Cocco11} is probably the best choice.
SCE, in fact, splits the system in many subunits that are recursively joined together to form larger and larger building blocks of the reconstructed interaction network.
If the interaction graph is clusterizable, SCE will recognizes these units and accomplish the inference task quickly.
However, up to our knowledge, a simple and generic argument for choosing the best suited algorithm for the actual inference problem is still missing and it would be of large interest.
 
\section*{Acknowledgments}  
U.F. thanks S. Deny, G. Gardella, O. Marre, R. Monasson, T. Mora, T. Obuchi and B. Telenczuk for useful discussions, S. Deny and O. Marre for the retina dataset and  A. Destexhe for hosting at the European Institute for Theoretical Neuroscience.
This research was supported by a grant from the Human Brain Project (HBP CLAP)

\appendix

\section{The existence and uniqueness of the inference solution}
\label{app::Existence}
Because the matrix $\mathcal{H}[{\bf X}]$, see eq. (\ref{hessian}),  measures the concavity of maximization problem (\ref{maximization}), its positiveness guarantees the existence and uniqueness of the inference solution ${\bf X^*}$.
As $\mathcal{H}[{\bf X}]$ can be expressed as the covariance matrix  $\chi[{\bf X}]$, see eq. (\ref{equalities}), it is by construction non-negative, but it could have zero modes.
A zero mode in a covariance matrix identifies a linear combination of the observables, the corresponding Eigenvector, that does not fluctuate within the model probability distribution. 
This could happens either for all the values of the parameter ${\bf X }$ when some of the functions  ${\bf \Sigma}({\bf \sigma})$ are linearly dependent either at the solution ${\bf X^*}$ when some of the fields diverge quenching (part of) the system\cite{Ackley85,Barton13}.
This last case usually happens when the dataset suffers of unsersampling. 

As an example, consider a dataset of two non constant spins $\sigma_1(b)$ and $\sigma_2(b)$ that within the dataset are never active together, so that $\sigma_1(b) \sigma_2(b) =0$ for $b=1,\dots,B$.
If we take only $\Sigma_1({\bf \sigma})=\sigma_1$ and $\Sigma_2({\bf \sigma})=\sigma_2$ as observables, the inference problem is well posed.
However if we include, for example, $\Sigma_3({\bf \sigma}) = \sigma_1 +\sigma_2$ or $\Sigma_4({\bf \sigma})=\sigma_1 \sigma_2$ the matrix $\chi[{\bf X^*}]$ will develop zero modes.
In the first case, because $\Sigma_3({\bf \sigma}) = \Sigma_1({\bf \sigma}) +\Sigma_2({\bf \sigma})$ in the latter case because $X_4 \to -\infty$ in order to fix $\langle   \Sigma_4({\bf \sigma}) \rangle_{\bf X^*}=0$.

\section{On the deterministic convergence of the Vanilla Gradient algorithm.}
\label{app::Approaching}

As introduced in sect. \ref{sect::Approaching},  for ${\bf X \approx X^*}$ the deterministic dynamics of the VG algorithm is exactly solvable upon projecting the fields on the $\chi[{\bf X^*}]$ Eigenvectors. 
Along a $\mu$-Eigenspace the convergence of the VG algorithm is not uniform and scales with the corresponding Eigenvalue $\l_\mu$ as $(1- \alpha \l_\mu)^t$.
Indeed by tuning $\a$ to speed up some particular direction, the others can suffers of very low convergence.
The learning rate that optimize the convergence speed along all direction simultaneously is
\begin{eqnarray}
\a^{\text{(VG)}}_{BEST} &\equiv& \arg \min_\a \Big[~ \max_\mu \big|1-\a \l_\mu \big| ~\Big] \\
&=&  \arg \min_\a \max \Big[~\big|1-\a \l_< \big| ~,~\big|1-\a \l_> \big|~\big] \nonumber
\end{eqnarray}
where  $\l_{>/<}$ are the largest/smallest Eigenvalue.
This equation can be solved by equating the two expression in the $\max$:
\begin{equation}
\big| 1- \a^{\text{(VG)}}_{BEST} \l_<\big| = \big| 1- \a^{\text{(VG)}}_{BEST} \l_> \big|
\end{equation}
and the solution reads:
\begin{equation}
\a^{\text{(VG)}}_{BEST} = \frac{2}{ \l_>+\l_<}~,
\label{alphaBest}
\end{equation}
In particular, $\a^{\text{(VG)}}_{BEST} $ can be squeezed to small value by large $\l_>$ preventing the learning along the direction with $\lambda_\mu \ll \l_>$.
As an example, for the biological data we will consider, see Fig. \ref{fig:fisherEv}, $\l_>/ \l_< \simeq 10^5$.
Moreover the ratio $\l_>/ \l_<$ has been shown to diverge in synthetic data of model at criticality \cite{Machta13}.

\section{The stationary distribution of the stochastic dynamics}
\label{app::StochasticDynamics}

As introduced  in sect. \ref{sect::StochasticDynamics}, the stochastic dynamics of the fields ${\bf X}$ is ruled by the discrete-time master equation:
\begin{equation}
P_{t+1}({\bf X'})  =  \int D{\bf X}~ P_t({\bf X}) ~W_{{\bf X}\to {\bf X'}}, \nonumber
\end{equation}
where the transition rates depend on the distribution of ${\bf \nabla }  l_{\bf X}^\text{MC}$. 

As ${\bf Q}_{\bf X}^\text{MC}$ are the average of the observables ${\bf \Sigma}$, for large $M$ we expect them to be almost Gaussian distributed with a covariance equal to the ${\bf \Sigma}$ covariance, namely $ {\bf \chi[X]}$,  divided by the number of MC measurements:
\begin{equation}
P\Big( {\bf Q}_{\bf X}^\text{MC} \Big) \approx \mathcal{N}\Big[ ~{\bf Q}[\bf X]~; ~ \frac{ {\bf \chi[X]} }{M} ~\Big]\Big( { \bf Q }_{\bf X}^\text{MC} \Big)~. \label{pOfQofX}
\end{equation}
Consequently, as ${\bf \nabla }  l_{\bf X}^\text{MC} \equiv B \big( \overline{\bf P} - {\bf Q}_{\bf X}^\text{MC} \big)$ we have:
\begin{equation}
P\big( {\bf \nabla } l_{\bf X}^\text{MC} \big) \approx \mathcal{N} \Big[~{\bf \nabla }  l[{\bf X}] ~; ~ \frac{ {\bf \chi[X]} }{M} ~\Big]\big( {\bf \nabla } l_{\bf X}^\text{MC}\big) \nonumber
\end{equation}
If the truncation (\ref{logLlongTime}) is valid and $\chi[{\bf X \approx X^*}] \approx \overline{{\bf~ \chi}~}$, we can replace the gradient mean by the derivative of the approximated log-likelihood,
\begin{eqnarray}
{\bf \nabla }  l[{\bf X}] &=& {\bf \nabla }  \big[-\frac{B}{2}  ({\bf X - X^*}) {\overline{{~ \chi}~}  }  ({\bf X - X^*})    \big] \label{approxGrad}\\
&=& B~ \overline{ ~ \bf\chi ~ }({\bf X^* - X})~,
\end{eqnarray}
to finally obtain:
\begin{equation}
P\big( {\bf \nabla } l_{\bf X}^\text{MC} \big) \approx \mathcal{N} \Big[B~ \overline{ ~ \bf\chi ~ }({\bf X^* - X}); \frac{B^2}{M}\overline{ \bf ~\chi ~ } \Big]\big( {\bf \nabla } l_{\bf X}^\text{MC}\big), \nonumber 
\label{pGradGivenX}
\end{equation}

The transition rates $W_{{\bf X}\to {\bf X'}}$ are the probability to measure a value of  $l_{\bf X}^\text{MC}$ such that the next fields value in the dynamics is ${\bf X'}$.
From eqs.  (\ref{VanillaGradient}) and (\ref{ApproximatedNewton}) it follows:
\begin{eqnarray}
W_{{\bf X}\to {\bf X'}}^\text{VG}  &=& \mathcal{N} \Big[ \overline{ ~ \bf\chi ~ }({\bf X^* - X}); \frac{\overline{ \bf ~\chi ~ }}{M} \Big]\Big( \frac{ {\bf X' - X} }{ \a }  \Big) \\
W_{{\bf X}\to {\bf X'}}^\text{DD} &=& \mathcal{N} \Big[ \overline{ ~ \bf\chi ~ }({\bf X^* - X}); \frac{\overline{ \bf ~\chi ~ }}{M} \Big]\Big( \frac{  \overline{~\bf \chi~} ({\bf X' - X})  }{\a } \Big)
\end{eqnarray}

By asking $P_t({\bf X})$ to be invariant under the evolution (\ref{masterEq}) we can obtain the stationary distribution $P_\infty({\bf X})$:
\begin{eqnarray}
P^\text{VG}_\infty( {\bf X} )   &=& \mathcal{N} \Big[ {\bf X^*} ; \frac{\a}{M } \big(2 {\bf \delta_D}- \alpha \overline{ \bf ~\chi ~ } \big)^{-1} \Big]( {\bf X} ),  \nonumber\\
P^\text{DD}_\infty( {\bf X} )   &=&  \mathcal{N} \Big[ {\bf X^*} ; \frac{\a}{M (2 - \alpha)} \overline{\bf ~\chi ~ }^{-1} \Big]( {\bf X} ),  \nonumber
\end{eqnarray}
where $\delta_D$ is the identity matrix in dimension $D$. Here the typical fluctuations of ${\bf X}$ around ${\bf X^*}$ must consistently verify the approximation ${\bf X \approx X^*}$: 
$\la {\bf (X - X^*)^2} \ra_{P_\infty }$ should be small enough to allow the expansion (\ref{logLlongTime}).

We can obtain conditions on the algorithm convergence by requiring $P_\infty(X)$ to be a properly defined probability distribution, namely to be integrable.
Remarkably by asking $P_\infty(X)$ to have a positive covariance, we re-obtain the upper-bounds for the learning rate $\a$: $\a \l_\mu < 2$, for all $\mu$, for the VG and $\a<2$ for the DD.

In the stationary regime, the fluctuating fields ${\bf X}$ will induce a second source of noise in the actual distribution of ${\bf Q}^\text{MC}$.
By inserting the approximation (\ref{approxGrad})  in the distribution (\ref{pOfQofX}) and then by averaging over $P_\infty({\bf X})$, we obtain:
\begin{eqnarray}
P^\text{VG}_\infty( {\bf Q}^\text{MC} ) &=& \mathcal{N} \Big[  \overline{\bf P}; \frac{2 \overline{\bf ~\chi~} \left(2 \d_D - \a \overline{ \bf ~\chi ~ } \right)^{-1}}{M} \Big]( {\bf Q}^\text{MC} ), \nonumber \\
P^\text{DD}_\infty( {\bf Q}^\text{MC} ) &=& \mathcal{N} \Big[  \overline{\bf P}; \frac{2  \overline{ \bf ~\chi ~ } }{M (2-\a)}\Big]( {\bf Q}^\text{MC} ) ~.\nonumber
\end{eqnarray}

$P_\infty({\bf X})$, see eq.s (\ref{PSofXVG}) and  (\ref{PSofXDD}), allows us to compute the expected deviation of the log-likelihood from $l[{\bf X^*}]$.
By averaging  $\delta l[{\bf X}]$, see eq. (\ref{logLlongTime}),  we obtain:
\begin{eqnarray}
\la \delta l \ra_{P^\text{DD}_\infty } &=& - \frac{D }{2} \frac{B~ \a}{M(2-\a)} \nonumber\\
 \sqrt{ \la \delta l^2 \ra_{P^\text{DD}_\infty } - \la \delta l \ra^2_{P^\text{DD}_\infty }  } &=&  \sqrt{\frac{D}{2}} \frac{B~ \a}{M (2-\a)}\nonumber
\end{eqnarray}
for the DD algorithm and
\begin{equation}
\la \delta l \ra_{P^\text{VG}_\infty } = - \frac{\a B}{2 M} \sum_\mu \frac{ \l_\mu }{2 - \a \l_\mu} \nonumber
\end{equation}
for the VG (we do not report the slightly involved expression of the variance, which also scales as $1/M$). 
Both estimations are indeed biased to lower values with large fluctuations (of the order of the bias itself).
For the DD algorithm the bias depends just on $D$, $\a$ and $M$ and consequently it is data independent. 
For the VG, instead, it depends strongly on the spectrum of $\overline{ \bf ~\chi ~ }$ and for large $\l_\text{MAX}$ or $\a \simeq  2/\l_\text{MAX}$ it can reach very large values, thus nullifying the inference effort. 

Note that the three matrix appearing in the covariance of eq. (\ref{pOfQofX}) and in the mean and covariance of eq. (\ref{pGradGivenX}) are \textit{a priori} different: the first is the model susceptibility, the second is the log-likelihood Hessian where the third is the model Fisher matrix.
However, as explained in section \ref{geometry} for the MaxEnt inference problem these three matrices coincide providing the results (\ref{PSofXVG}) and (\ref{PSofXDD}).

\section{The posterior sampling avoids to over-estimate the log-likelihood}
\label{logL_comparison}

To better understand the consequences of over-fitting we consider now the case where the system that generates the data is of MaxEnt form with some unknown true fields ${\bf \hat X}$.
By ideally sampling the distribution infinitely many times we can access to the true means of the conjugated observable ${\bf \hat P}$ and from these compute the true log-likelihood:
\begin{equation}
\wh{l} = {\bf \hat P} \cdot {\bf \hat X} - \ln Z[ {\bf \hat X}]~.
\end{equation}
We like to compare the true log-likelihood with that obtained by the exact inference, the one leading to ${\bf X^*}$  and with that obtained by the DD algorithm which samples from the posterior.
By sampling $B$ times from the true model distribution we can generate synthetic dataset and obtain \textit{empirical} estimates $\overline{ \bf P}$ of ${\bf \hat P}$.
Through the central limit theorem we can approximate the distribution of the  expected deviation of the observable means:
\begin{equation}
P\big(~ { \bf \delta P}~ \big) \equiv P\big(~ \overline{ \bf P} - \hat{\bf P} ~\big) = \mathcal{N} \Big[ ~0~ ;~ \frac{ \hat \chi}{B} ~\Big](  \overline{ \bf P} )\label{pOfDeltaP}
\end{equation}
where $\hat \chi = \chi[{\bf \hat X}]$ is the susceptibility matrix of true model.

The exact inference of the fields perfectly reproducing $\overline{ \bf P}$ overestimates the log-likelihood, in fact:
\begin{eqnarray}
l^*_{{\bf \delta P}} &=& \max_{\bf X} \Big[ \big({\bf \hat P +\delta P}\big) \cdot {\bf X} - \ln Z[ {\bf X}]  \Big] \nonumber \\
&=& \max_{\bf \delta X} \Big[ \big( {\bf \hat P +\delta P}\big) \cdot \big( {\bf \hat X + \delta X}\big) - \ln Z[ {\bf \hat X + \delta X}]  \Big] \nonumber \\
&\approx& \wh{l} + {\bf \delta P}\cdot {\bf \hat X} +  \max_{\bf \delta X}\Big[   {\bf \delta P} \cdot {\bf \delta X}  - \frac{1}{2} {\bf \delta X} \cdot \hat \chi  \cdot {\bf \delta X} \Big] \nonumber  \\
&=&  \wh{l} + {\bf \delta P}\cdot {\bf \hat X} + \frac{1}{2} {\bf \delta P} \cdot \hat \chi^{-1}  \cdot {\bf \delta P} \label{logLstar}
\end{eqnarray}
where we approximate $\ln Z[ {\bf \hat X + \delta X}]$ up to the second order.
Through the expression (\ref{pOfDeltaP}) we can approximate the distribution of $l^*_{{\bf \delta P}}$ over many realization of the synthetic experiment:
\begin{equation}
P\big(~l^* ~ \big)  = \mathcal{N} \Big[ ~\wh{l} + \frac{D}{2 B} ~ ;~    \frac{ {\bf \hat X}  \cdot \hat \chi \cdot {\bf \hat X}  }{B} ~\Big]( l^* )\label{pLogLstar}
\end{equation}
where as before, $D$ is the dimension of the fields vector and we discard terms of order $B^{-2}$. 
$l^*$ is on average positively biased by a factor $\frac{D}{2 B}$.

In the case of the DD algorithm ($\a=1$ and $M=B$)  we have to substitute the maximization over ${\bf X}$ with an integration over the stationary fields distribution (\ref{PSofXDD}), which in this case will read:
\begin{eqnarray}
P^\text{DD}_\infty(~ {\bf \delta X}~ |~ {\bf \delta P}~ )   &\equiv& P^\text{DD}_\infty(~ {\bf X - \hat X}~ |~ {\bf \delta P}~ ) \nonumber \\ 
&=& \mathcal{N} \Big[~ \hat \chi^{-1}  \cdot {\bf \delta P} ~ ;~ \frac{ {\bf \hat \chi ~ }^{-1}}{B}~ \Big]( {\bf \delta  X} )
\end{eqnarray}
where the mean equals value of ${\bf \delta X}$ after the maximization in the calculation of $l^*_{{\bf \delta P}}$, see eq. (\ref{logLstar}).
For the DD algorithm we obtain:
\begin{eqnarray}
l^\text{DD}_{{\bf \delta P}} &\approx& \wh{l} + {\bf \delta P}\cdot {\bf \hat X} +  \Big\langle   {\bf \delta P} \cdot {\bf \delta X}  - \frac{1}{2} {\bf \delta X} \cdot \hat \chi  \cdot {\bf \delta X} \Big\rangle_{P^\text{DD}_\infty}  \nonumber  \\
&=&  \wh{l} + {\bf \delta P}\cdot {\bf \hat X} + \frac{1}{2} {\bf \delta P} \cdot \hat \chi^{-1}  \cdot {\bf \delta P} -\frac{D}{2 B} \nonumber \\
&=& l^*_{{\bf \delta P}} -\frac{D}{2 B} \label{logLdd}
\end{eqnarray}
and again through (\ref{pOfDeltaP}) we obtain:
\begin{equation}
P\big(~l^\text{DD} ~ \big)  = \mathcal{N} \Big[ ~\wh{l} ~ ;~    \frac{ {\bf \hat X}  \cdot \hat \chi \cdot {\bf \hat X}  }{B} ~\Big]( l^\text{DD} )\label{pLogLdd}~,
\end{equation}
where again we discard terms of order $B^{-2}$.
Coherently, the integration over the posterior distribution, that prevents to exactly maximize the log-likelihood, cancels the bias in the average of (\ref{pLogLdd}).

\begin{figure}[t]
\includegraphics[clip=true,keepaspectratio,angle=-0,width=\columnwidth]{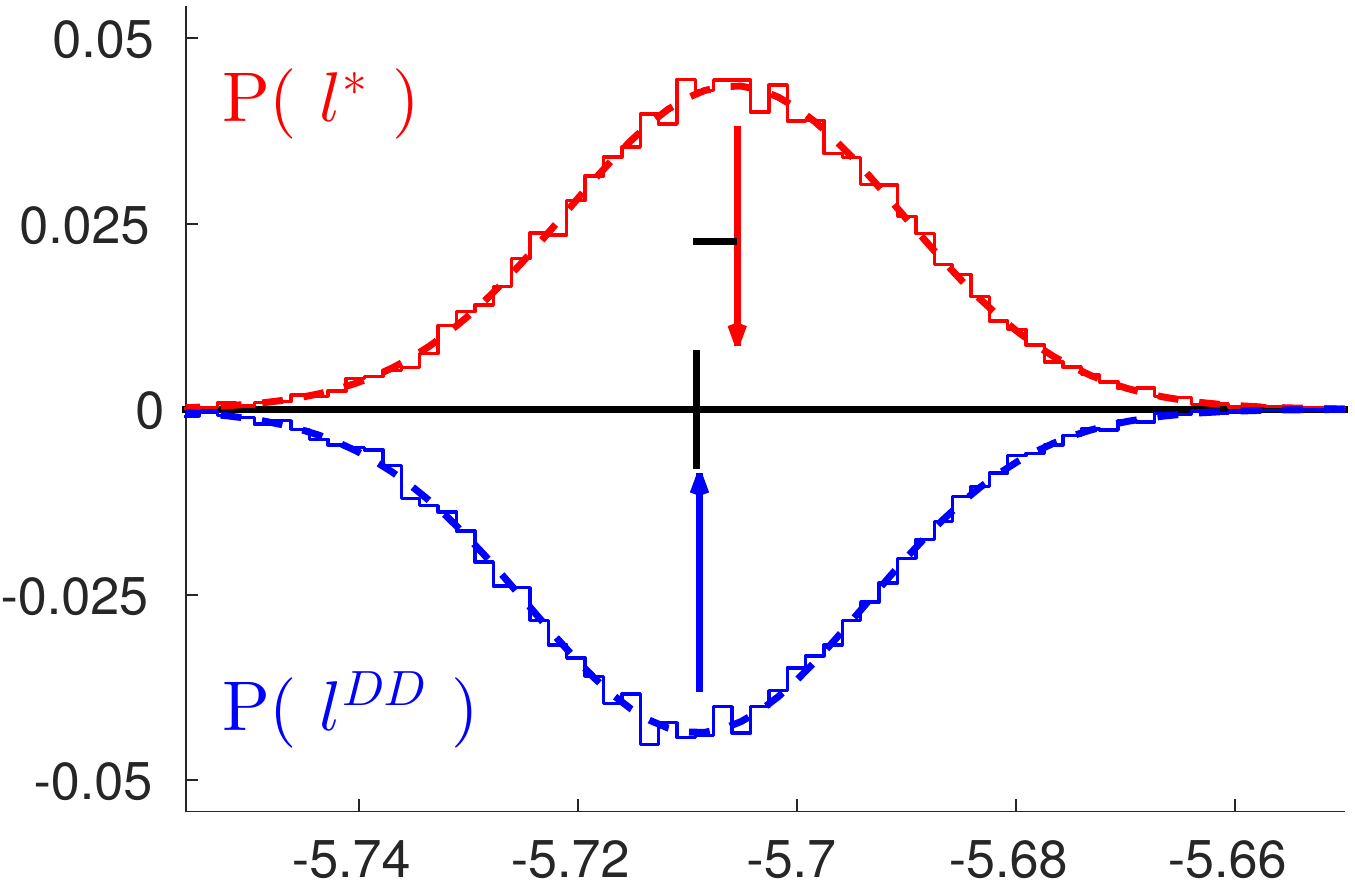}
\caption{(Colors online).
Histogram of  $l^*_{{\bf \delta P}}$ (straight) and $l^\text{DD}_{{\bf \delta P}} $ (reflected) obtained from the inference of $S=2 \cdot 10^4$ \textit{empirical} estimation of $\overline{ \bf\delta  P}$.
See text for the details of the synthetic pairwise Ising model used for the simulations.
Arrows indicate the corresponding histograms empirical means. 
The black vertical bar indicates the numerically exact value of $ \wh{l}$, the exact model log-likelihood, whereas the black horizontal segment measures $D/(2B)$. 
Dotted lines represents the theoretical distributions, see eqs. (\ref{pLogLstar}) and (\ref{pLogLdd}).
As predicted, see text, the average of $l^\text{DD}_{{\bf \delta P}}$ coincides with $\wh{l}$, whereas the average of $l^*_{{\bf \delta P}}$ suffers by a bias of $D/(2B)$
}
\label{fig:logL_comp}
\end{figure}

To test these results we perform an analysis on a pairwise Ising model of $N=10$ units, thus restricting the observables list to single and pairwise correlations ($\{ \Sigma_a\}_{a=1}^D = \{ \{\s_i\}_{i=1}^N ,  \{\s_i \s_j\}_{i<j=1}^N\}$) and consistently the fields to biases and pairwise interactions ($\{ X_a\}_{a=1}^D = \{ \{h_i\}_{i=1}^N ,  \{J_{ij}\}_{i<j=1}^N\}$). 
As synthetic model we chose a diluted disorder Ising model on a Erd\H{o}s-R\'enyi random graph with average connectivity $c=4$:
\begin{equation}
\begin{array}{c c c c c}
 J_{ij} & = & ~1.33 & \quad, \qquad & \text{with prob.} \frac{c}{2(N-1)} \\
 J_{ij} & = & -1.33 & \quad , \qquad& \text{with prob.} \frac{c}{2(N-1)} \\
J_{ij}  & = &  ~0,   & \quad, \qquad & \text{otherwise}\\
h_i    & = & - 0.46  & + \sum_{j \neq i} J_{ij} &
\end{array}
\end{equation}

We first computed $ \wh{l} = 5.7092$ and then we simulated the model  $S=2\cdot 10^4$ times to collect $S$ \textit{empirical} estimates of $ \overline{ \bf\delta  P}$.
The number of MCMC sampled was fixed at $B=2^{13}$.
For each of these $S$ realization we estimate $l^*_{{\bf \delta P}}$ and $l^\text{DD}_{{\bf \delta P}} $ and in Fig. \ref{fig:logL_comp} we compare their histograms.
Dotted lines correspond to the theoretical distributions (\ref{pLogLstar}) and (\ref{pLogLdd}), whereas the arrows indicate the empirical means.
The black vertical lines represents $ \wh{l}$. 
As can be appreciated by the small difference between the black vertical bar and the blue arrow, the average over the posterior distribution distribution removes the bias produced by the exact inference.
Moreover the difference between the red and blue arrows is approximately $D/(2B)$, equals to the length of the horizontal segment. 
More precisely:
\begin{eqnarray}
\wh{l} -  \Big\langle l^*_{{\bf \delta P}}  \Big\rangle_{P(~ {\bf \delta P}~)}  &=& (-35 ~ \!\pm 3.0) \cdot10^{-4}   \nonumber  \\
\frac{D}{2 B} + \wh{l} -  \Big\langle l^*_{{\bf \delta P}}  \Big\rangle_{P(~ {\bf \delta P}~)}   &=& (-1.2 \pm 3.0) \cdot10^{-4} \nonumber   \\
\wh{l} -  \Big\langle l^\text{DD}_{{\bf \delta P}}  \Big\rangle_{P(~ {\bf \delta P}~)} &=& (-0.3 \pm 0.6) \cdot10^{-4} \nonumber  
\end{eqnarray}

\bibliographystyle{unsrt} 
\bibliography{Ulisse.bib}

\begin{thebibliography}{10}

\bibitem{Buzaki04}
G.~Buz\'aki.
\newblock \em{Large-scale recording of neuronal ensembles }\em.
\newblock {\em \em Nat Neurosci. \em}, {\bf 7(5)}:446--51, 2004.

\bibitem{Phillips04}
S.~J. Phillips, M.~Dudík, and R.E. Schapire.
\newblock \em{A maximum entropy approach to species distribution modeling}\em.
\newblock {\em \em In Proceedings of the Twenty-First International Conference
  on Machine Learning \em}, 2004.

\bibitem{Schneidman06}
E.~Schneidman, M.~Berry, R.~Segev, and W.~Bialek.
\newblock \em{Weak pairwise correlations imply strongly correlated network
  states in a population }\em.
\newblock {\em \em Nature \em}, {\bf 440}:1007, 2006.

\bibitem{Peyrache09}
A.~Peyrache, M.~Khamassi, K.~Benchenane, S.I. Wiener, and F.P. Battaglia.
\newblock \em{Replay of rule-learning related neural patterns in the prefrontal
  cortex during sleep }\em.
\newblock {\em \em Nat. Neurosci. \em}, {\bf 12 }:919--26, 2009.

\bibitem{Weigt09}
M.~Weigt, R.A. White, H.~Szurmant, J.A. Hoch, and T.~Hwa.
\newblock \em{ Identification of direct residue contacts in protein–protein
  interaction by message passing }\em.
\newblock {\em \em PNAS\em}, {\bf 106(1)}:67--72, 2009.

\bibitem{Cocco13}
S.~Cocco, R.~Monasson, and M.~Weigt.
\newblock \em{From principal component to direct coupling analysis of
  coevolution in proteins: Low-eigenvalue modes are needed for structure
  prediction}\em.
\newblock {\em \em PLoS Comput Biol \em}, {\bf 9}:E1003176, 2013.

\bibitem{Bialek14}
W.~Bialek, A.~Cavagna, I.~Giardina, T.~Mora, O.~Pohl, E.~Silvestri, M.~Viale,
  and A.~Walczak.
\newblock \em{Social interactions dominate speed control in driving natural
  flocks toward criticality.}\em.
\newblock {\em \em PNAS\em}, {\bf 111(20)}:7212--7217, 2014.

\bibitem{Santolini14}
M.~Santolini, T.~Mora, and V.~Hakim.
\newblock \em{A General Pairwise Interaction Model Provides an Accurate
  Description of In Vivo Transcription Factor Binding Sites. }\em.
\newblock {\em \em PLoS Comput Biol \em}, {\bf 9(6)}:E99015, 2014.

\bibitem{Jaynes82}
E.~T. Jaynes.
\newblock \em{ On The Rationale of Maximum-Entropy Method}\em.
\newblock {\em \em Proc. IEEE \em}, {\bf 70}:939, 1982.

\bibitem{Cocco09}
S.~Cocco, S.~Leibler, and R.~Monasson.
\newblock \em{ Neuronal couplings between retinal ganglion cells inferred by
  efficient inverse statistical physics methods }\em.
\newblock {\em \em Proc. Natl. Acad. Sci. USA \em}, {\bf 106}:14058, 2009.

\bibitem{Hamilton13}
L.~S. Hamilton, J.~Sohl-Dickstein, A.~G. Huth, V.~M. Carels, K.~Deisseroth, and
  S.~Bao.
\newblock \em{Optogenetic Activation of an Inhibitory Network Enhances
  Feedforward Functional Connectivity in Auditory Cortex}\em.
\newblock {\em \em Neuron \em}, {\bf 80}:1066--76, 2013.

\bibitem{Mora15}
T.~Mora, S~Deny, and O~Marre.
\newblock \em{Dynamical criticality in the collective activity of a population
  of retinal neurons }\em.
\newblock {\em \em Phys. Rev. Lett. \em}, {\bf 114}:078105, 2015.

\bibitem{Tavoni14}
T.~Tavoni, U.~Ferrari, S.~Cocco, F.P. Battaglia, and R.~Monasson.
\newblock \em{Inferred network of the prefrontal cortex activity unveils
  task-related coupling potentiations and cell assemblies }\em.
\newblock {\em \em submitted \em}, {\bf }, 2014.

\bibitem{Lezon06}
T.~R. Lezon, J.~R. Banavar, M.~Cieplak, A.~Maritan, and N.~V. Fedoroff.
\newblock \em{Using the principle of entropy maximization to infer genetic
  interaction networks from gene expression patterns}\em.
\newblock {\em \em PNAS\em}, {\bf 103}:19033--19038, 2006.

\bibitem{Ferguson13}
A.L. Ferguson, J.K. Mann, S.~Omarjee, T.~Ndung'u, B.D. Walker, and
  A.~Chakraborty.
\newblock \em{Translating HIV sequences into quantitative fitness landscapes
  predicts viral vulnerabilities for rational immunogen design.}\em.
\newblock {\em \em Immunity \em}, {\bf 38}:606--617, 2013.

\bibitem{Mann14}
J.K. Mann, J.P. Barton, A.L. Ferguson, S.~Omarjee, B.D. Walker, Chakraborty A.,
  and T.~Ndung'u.
\newblock \em{The Fitness Landscape of HIV-1 Gag: Advanced Modeling Approaches
  and Validation of Model Predictions by In Vitro Testing}\em.
\newblock {\em \em PLoS Comput Biol \em}, {\bf 10(8)}:e1003776, 2014.

\bibitem{Ganmor11}
E.~Ganmor, R.~Segev, and E.~Schneidman.
\newblock \em{Sparse low-order interaction network underlies a highly
  correlated and learnable neural population code }\em.
\newblock {\em \em PNAS\em}, {\bf 108}:9679--9684, 2011.

\bibitem{Tkacik14}
G.~Tkacik, O.~Marre, D.~Amodei, E.~Schneidman, W~Bialek, and Berry M.J.
\newblock \em{Searching for collective behaviour in a network of real neurons
  }\em.
\newblock {\em \em PloS Comput. Biol.\em}, {\bf 10(1)}:e1003408, 2014.

\bibitem{Ackley85}
D.~H. Ackley, G.~E. Hinton, and T.~J. Sejnowski.
\newblock A learning algorithm for boltzmann machines.
\newblock {\em Cognitive Science}, 9:147--169, 1985.

\bibitem{Hinton02}
G.~Hinton.
\newblock \em{Training Products of Experts by Minimizing Contrastive
  Divergence}\em.
\newblock {\em \em Neural Comput. \em}, {\bf 14(8)}:1771--1800, 2002.

\bibitem{Broderik07}
T.~Broderick, M.~Dudik, G.~Tkacik, R.E. Schapire, and W.~Bialek.
\newblock \em{Faster solutions to the inverse pairwise Ising problem }\em.
\newblock {\em Arxiv:0712.2437}, 2007.

\bibitem{Cocco11}
S.~Cocco and R.~Monasson.
\newblock \em{Adaptive cluster expansion for inferring Boltzmann machines with
  noisy data}\em.
\newblock {\em \em Phys. Rev. Lett. \em}, {\bf 106}:090601, 2011.

\bibitem{Barton13}
J.~Barton and S.~Cocco.
\newblock \em{Ising models for neural activity inferred via selective cluster
  expansion: structural and coding properties }\em.
\newblock {\em \em J. Stat. Mech. \em}, {\bf}:P03002, 2013.

\bibitem{Sohl-Dickstein11}
J.~Sohl-Dickstein, P.~B. Battaglino, and M.~R. DeWeese.
\newblock \em{New Method for Parameter Estimation in Probabilistic Models:
  Minimum Probability Flow}\em.
\newblock {\em \em Phys. Rev. Lett. \em}, {\bf 107}:220601, 2011.

\bibitem{Kappen97}
H.J. Kappen and F.B. Rodriguez.
\newblock \em{Efficient learning in boltzmann machines using linear response
  theory.}\em.
\newblock {\em \em Neural Comput. \em}, {\bf 10}:1137--1156, 1997.

\bibitem{Tanaka98}
T.~Tanaka.
\newblock \em{Mean-field theory of Boltzmann machine learning}\em.
\newblock {\em \em Phys. Rev. E\em}, {\bf 58}:2302, 1998.

\bibitem{Aurell12}
E~Aurell and M.~Ekeberg.
\newblock \em{Inverse Ising Inference Using All the Data }\em.
\newblock {\em \em Phys. Rev. Lett. \em}, {\bf 108}:090201, 2012.

\bibitem{Ricci12}
F~Ricci-Tersenghi.
\newblock \em{ The Bethe approximation for solving the inverse Ising problem: a
  comparison with other inference methods }\em.
\newblock {\em \em J. Stat Mech \em}, page P08015, 2012.

\bibitem{Jacquin15}
H.~Jacquin and A.~Rancon.
\newblock \em{Efficient, fast and principled mean-field inference for strongly
  coupled data}, {2015}.

\bibitem{Note1}
In order to lighten the notation we do not distinguish between column or row
  vector and we avoid any transpose symbol.

\bibitem{Amari98}
S.~Amari.
\newblock \em{Natural Gradient Works Efficiently in Learning, Neural
  Computation}\em.
\newblock {\em \em Neural Comput. \em}, {\bf 10}:251--276, 1998.

\bibitem{Press07}
W.H. Press, S.~A. Teukolsky, V.~T. Vetterling, and B.~P. Flannery.
\newblock {\em Numerical Recepies}.
\newblock Cambridge University Press (Cambridge, U.K.), 2007.

\bibitem{Amari98b}
S.~Amari and S.C. Douglas.
\newblock \em{ Why natural gradient?}\em.
\newblock {\em \em Proc. IEEE \em}, {\bf 2}:1213--16, 1998.

\bibitem{Amari07}
S.~Amari and H.~Nagaoka.
\newblock \em{Methods of information geometry}\em.
\newblock Oxford University Press, Oxford, 2007.

\bibitem{Note2}
As the matrix $\protect \overline { ~ \protect \bf \chi ~ }$ has several
  symmetries a compressed encoding could decrease the required storage memory.

\bibitem{Machta13}
B.B. B.~Machta, R.~Chachra, M.K. Transtrum, and J.P. Sethna.
\newblock \em{Parameter Space Compression Underlies Emergent Theories and
  Predictive Models}\em.
\newblock {\em \em Science \em}, {\bf 342}:604--607, 2013.

\bibitem{Marre12}
O.~Marre, D.~Amodei, N.~Deshmukh, K.~Sadeghi, F.~Soo, T.~Holy, and M.J. Berry.
\newblock \em{Recording of a large and complete population in the retina}\em.
\newblock {\em \em Journal of Neuroscience \em}, {\bf 32(43)}:1485973, 2012.

\bibitem{Peyrache12}
A.~Peyrache, N.~Dehghani, Eskandar~E. N., J.~R. Madsen, W.~S. Anderson, L.R.
  Donoghue, J.A.~Hochberg, E.~Halgren, S.S. Cash, and A.~Destexhe.
\newblock \em{Spatiotemporal dynamics of neocortical excitation and inhibition
  during human sleep }\em.
\newblock {\em \em Pnas \em}, {\bf 109 }:1731--36, 2012.

\bibitem{Welling11}
M.~Welling and Y.W. Teh.
\newblock Bayesian learning via stochastic gradient langevin dynamics.
\newblock {\em Proceedings of the 28th International Conference on Machine
  Learning (ICML)}, page 681–688, 2011.

\end{thebibliography}


\end{document}